\documentclass[preprinty]{aastex}
\usepackage{longtable}
\usepackage{graphicx,amssymb,mathrsfs,amsmath}
\usepackage{lscape}

\begin{document}

\title{THE KINEMATICS AND CHEMISTRY OF RED HORIZONTAL BRANCH\\
       STARS IN THE SAGITTARIUS STREAMS}

\author{W.B.SHI\altaffilmark{1,2}
        Y.Q.CHEN\altaffilmark{1}
        K.CARRELL\altaffilmark{1} and 
        G.ZHAO\altaffilmark{1,2}$^{\dag}$}

\affil{1 Key Laboratory of Optical Astronomy, National Astronomical
  Observatories, Chinese Academy of Sciences, Beijing 100012, China;
 swb@sdu.edu.cn; cyq@bao.ac.cn; carrell@nao.cas.cn; gzhao@nao.cas.cn}

\affil{2 Shandong Provincial Key Laboratory of Optical Astronomy and
  Solar-Terrestrial Environment, School of Space Science and Physics,
  Shandong University at Weihai, Weihai 264209, China}

\begin{abstract}
  We have selelcted 556 Red Horizontal Branch (RHB) stars along the
  streams of the Sagittarius dwarf galaxy (Sgr) from SDSS DR7
  spectroscopic data using a theoretical model. The metallicity and
  $\alpha$-elements distributions are investigated for stars in the
  Sgr streams and for Galactic stars at the same locations. We find
  that the Sgr stars have two peaks in the metallicity distribution
  while the Galactic stars have a more prominent metal-poor peak.
 Meanwhile, [$\alpha$/Fe]
  ratios of the Sgr stars are lower than those of the Galactic
  stars. Among the Sgr stars, we find a difference in the metallicity
  distribution between the leading and trailing arms of the Sgr tidal
  tails. The metallicity and [$\alpha$/Fe] distribution of the leading
  arm is similar to that of the Galaxy. The trailing arm is composed
  mainly of a metal rich component and [$\alpha$/Fe] is obviously
  lower than that of the Galactic stars. The metallicity gradient is
  -(1.8 $\pm$ 0.3)$\times10^{-3}$ dex degree$^{-1}$ in the first wrap
  of the trailing arm and -(1.5 $\pm$ 0.4)$\times10^{-3}$ dex
  degree$^{-1}$ in the first wrap of the leading arm. No significant
  gradient exists along the second wraps of the leading or trailing
  arms. It seems that the Sgr dwarf galaxy initially lost the metal
  poor component in the second wrap (older) arms due to the tidal
  force of our Galaxy and then the metal rich component is disrupted
  in the first wrap (younger) arms. Finally, we found that the
  velocity dispersion of the trailing arm from
  $88^\circ<\Lambda_\odot<112^\circ$ is $\sigma$ = 9.808 $\pm$
  1.0~km~s$^{-1}$, which is consistent with previous work in the
  literature.
\end{abstract}


\section{Introduction}

The Sagittarius dwarf galaxy is the second nearest galaxy to our Milky
Way(assuming the Canis Major dwarf galaxy is the nearest). The Sgr is currently being
disrupted under the strain of the Milky Way. Studying the metallicity
and kinematic distributions of Sgr stars has now become an important
issue. Many works on chemical abundances of the Sgr stars have been
done based on high resolution spectra. \citet{bel08} selected 321 RGB
stars in the Sgr nucleus and give the average [Fe/H]$\sim$ -0.45\,dex
from the infrared Ca II triplet. \citet{car10} derived homogeneous
elemental abundances with 27 red giant stars belonging to the Sgr
nucleus and found on average [Fe/H] $\sim$ -0.61\,dex --
-0.74\,dex. \citet{kel10} observed 11 M giant stars with the Gemini
South telescope which indicated the [Fe/H] of stars decreases along
the tidal stream. \citet{cho07} present a reliable measurement on M
giants with high resolution at different points along the tidal stream
and show a significant metallicity gradient. They found a median
[Fe/H]$\sim$-0.4 in the core that decreases to -1.1 dex over the
leading arm.

However, these works based on high resolution spectra have small
samples of stars. Based on low resolution spectra \citet{yan09} traced
the Sgr tidal streams with red K/M-giants from the SDSS survey. They
found an average [Fe/H] in the range -0.8$\pm$0.2 with 33 K/M-giant 
stars in two areas. \citet{carlin12} derived metallicity from
low-resolution spectra of stars along a stretch of the Sgr stream and
find a constant [Fe/H]$\sim$-1.15. Thus far, the metallicity and
abundance studies of the Sgr tails have been less detailed in large
samples of stars and in various locations, which are the advantages of
the present work. We analyze the metallicity distribution at different
points along the tidal streams of the Sgr with low resolution data for
a large sample of stars.

The SDSS spectroscopic survey and the LAMOST project \citep{zha06},
will provide a large sample of Red Horizontal Branch (RHB) stars with
low resolution spectra in the Sgr. Currently, the spectroscopic data
of the SDSS survey provides stellar parameters, distances, radial
velocities and metallicities for many stars spread across a wide
area. In this work, we investigate the properties of Sgr RHB stars
from SDSS and compare them with stars in the Milky Way. We present the
procedure for selecting the RHB stars in Sect.2 and give the
metallicity analysis in Sect.3. In Sect.4 we test the theoretical
model and sample selection, and a summary is given in Sect.5.

\section{Sample selection}
We obtained 8535 RHB Stars, 5391 with (U,V,W) and 3144 stars without,
from SDSS DR7 low resolution spectral data \citep{che10}. We choose
the Sgr stars with the aid of a theoretical model by
\citet{law10}. \citet{law10} provide a model of the Sgr orbiting in a
triaxial Galactic potential with $10^5$ points. The model divides the
$10^5$ points into four parts: leading arm 1 and 2 and trailing arm 1
and 2. The model is based on observational data from 2MASS and SDSS
(for more details please see \citet{law10}).

\begin{figure*}[!htbp]
\begin{minipage}[t]{0.5\textwidth}
\centering
{\bf Leading arm 1}
\end{minipage}
\begin{minipage}[t]{0.5\textwidth}
\centering
{\bf Leading arm 2}
\end{minipage}
\begin{minipage}{\textwidth}
\includegraphics[width=0.35\textwidth,angle=-90,clip]{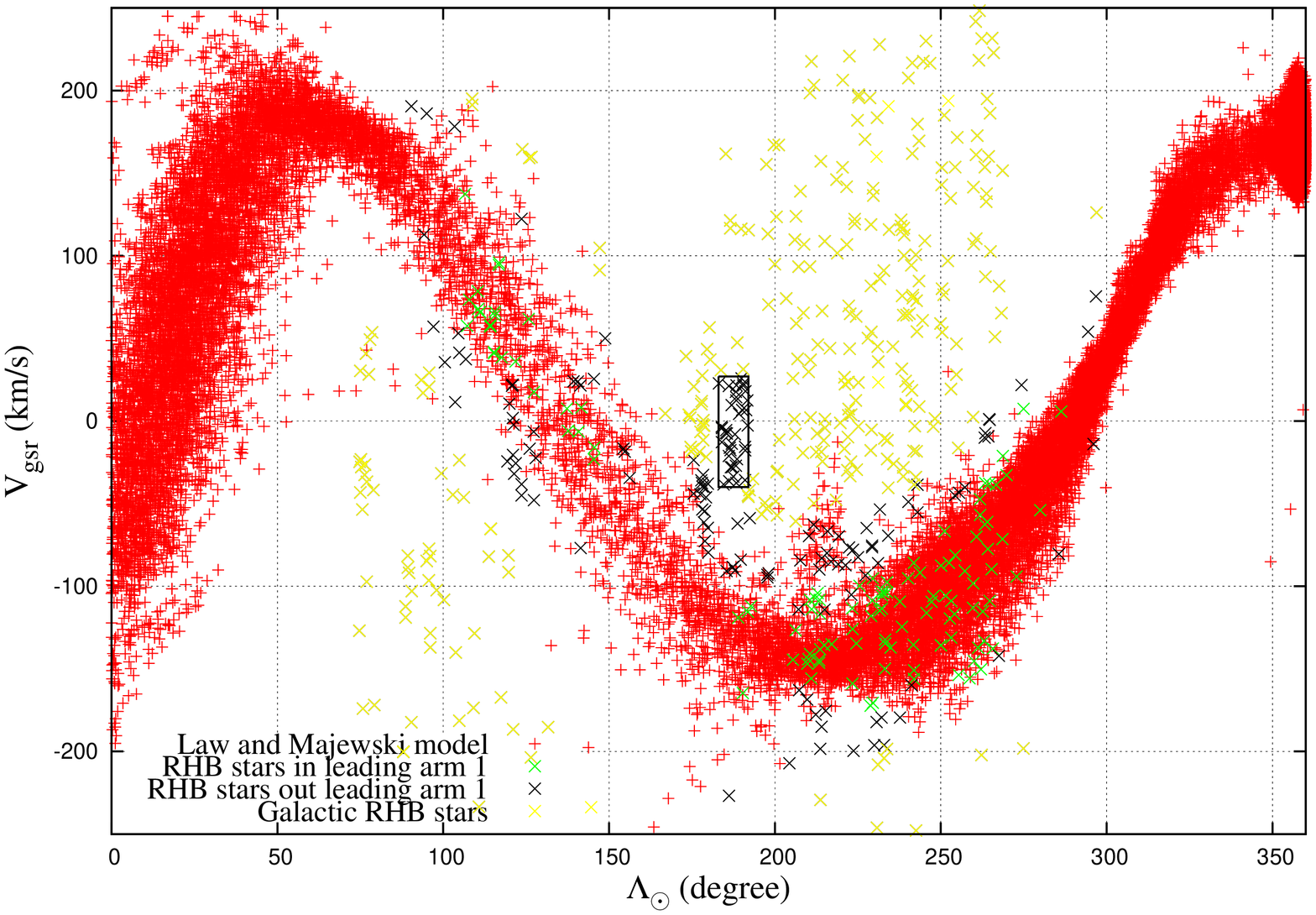}
\hspace*{\fill}
\includegraphics[width=0.35\textwidth,angle=-90,clip]{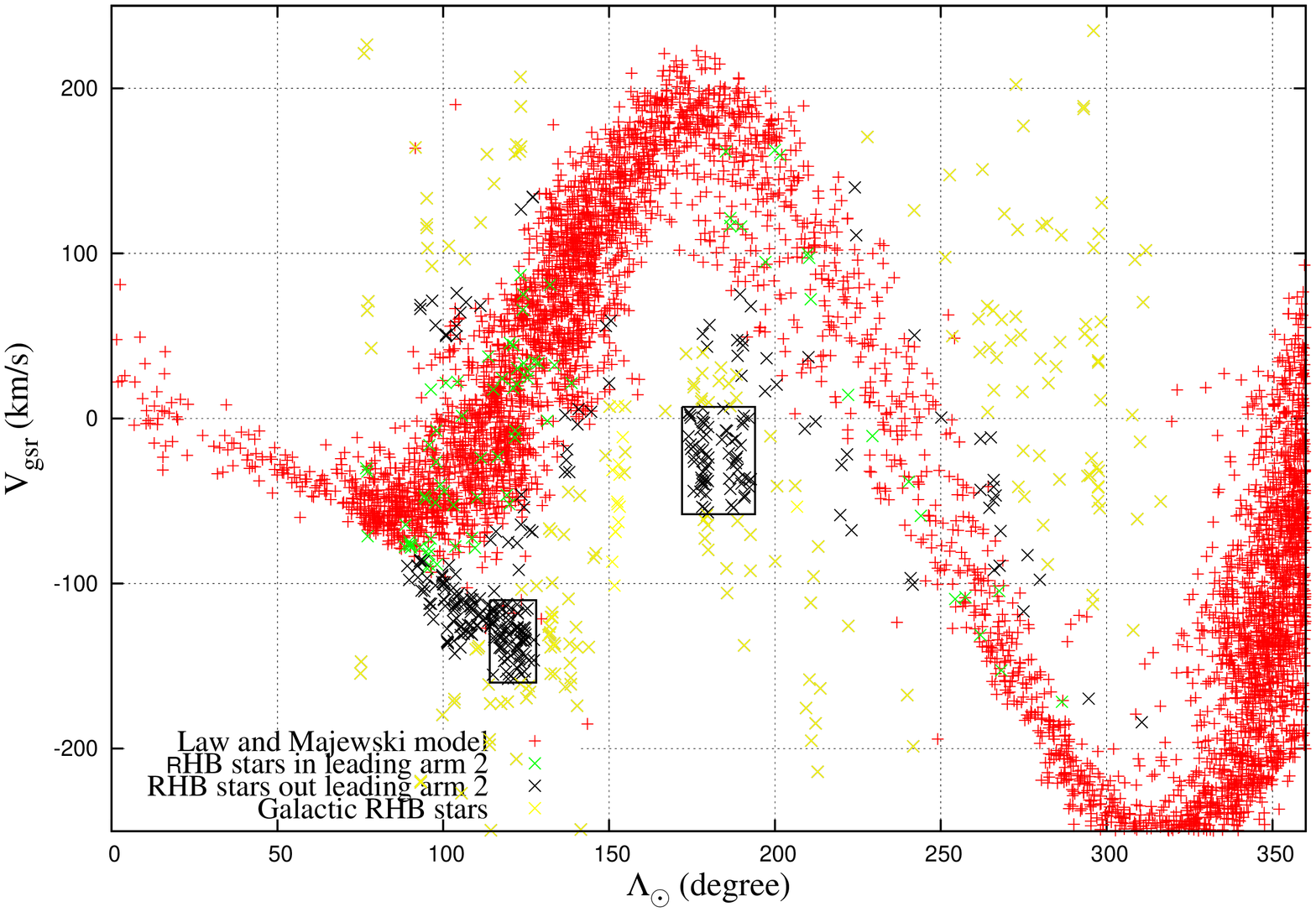}
\end{minipage}\vspace{0.2cm}
\begin{minipage}[t]{0.5\textwidth}
\centering
{\bf Trailing arm 1}
\end{minipage}
\begin{minipage}[t]{0.5\textwidth}
\centering
{\bf Trailing arm 2}
\end{minipage}
\begin{minipage}{\textwidth}
\includegraphics[width=0.35\textwidth,angle=-90,clip]{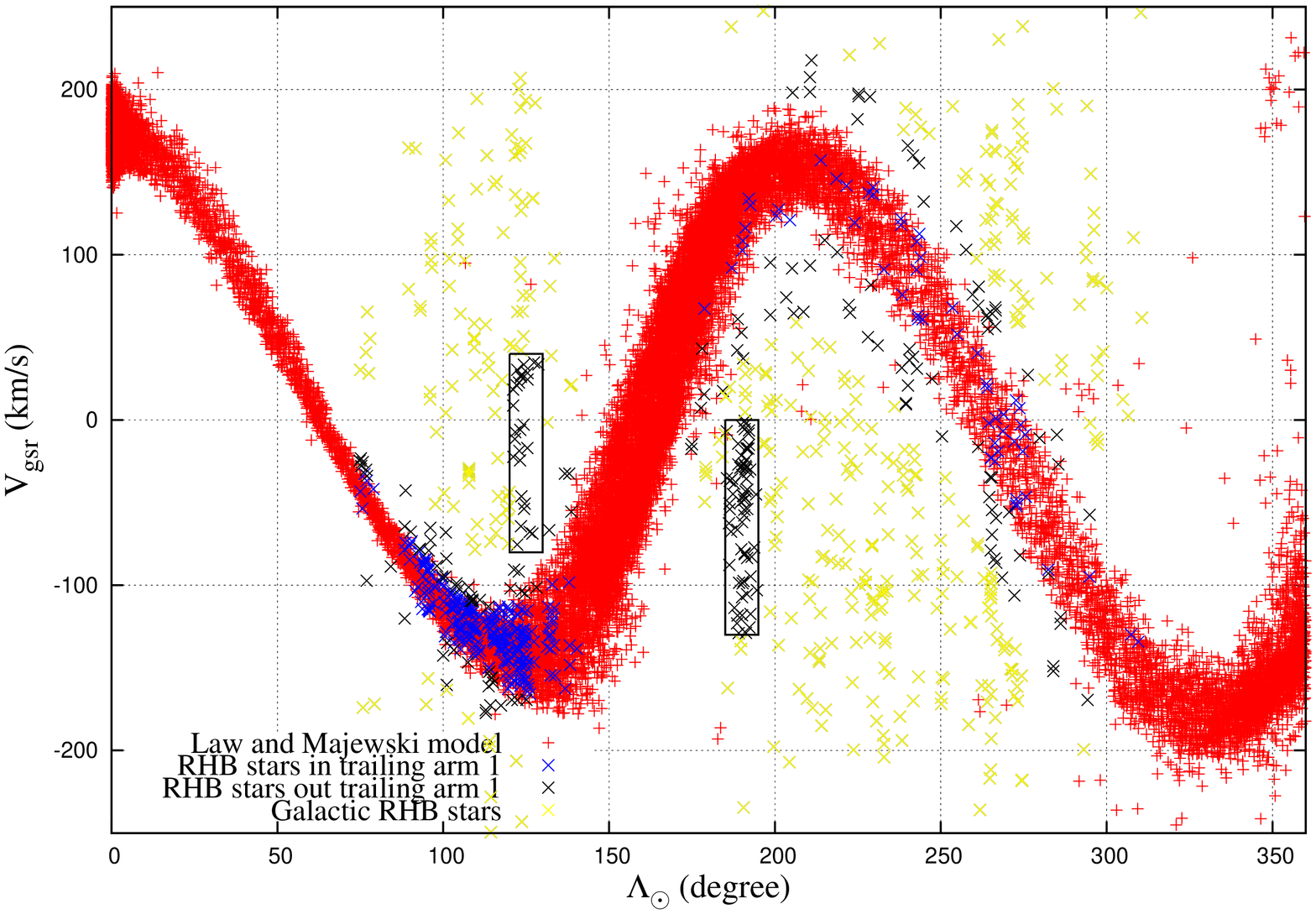}
\hspace*{\fill}
\includegraphics[width=0.35\textwidth,angle=-90,clip]{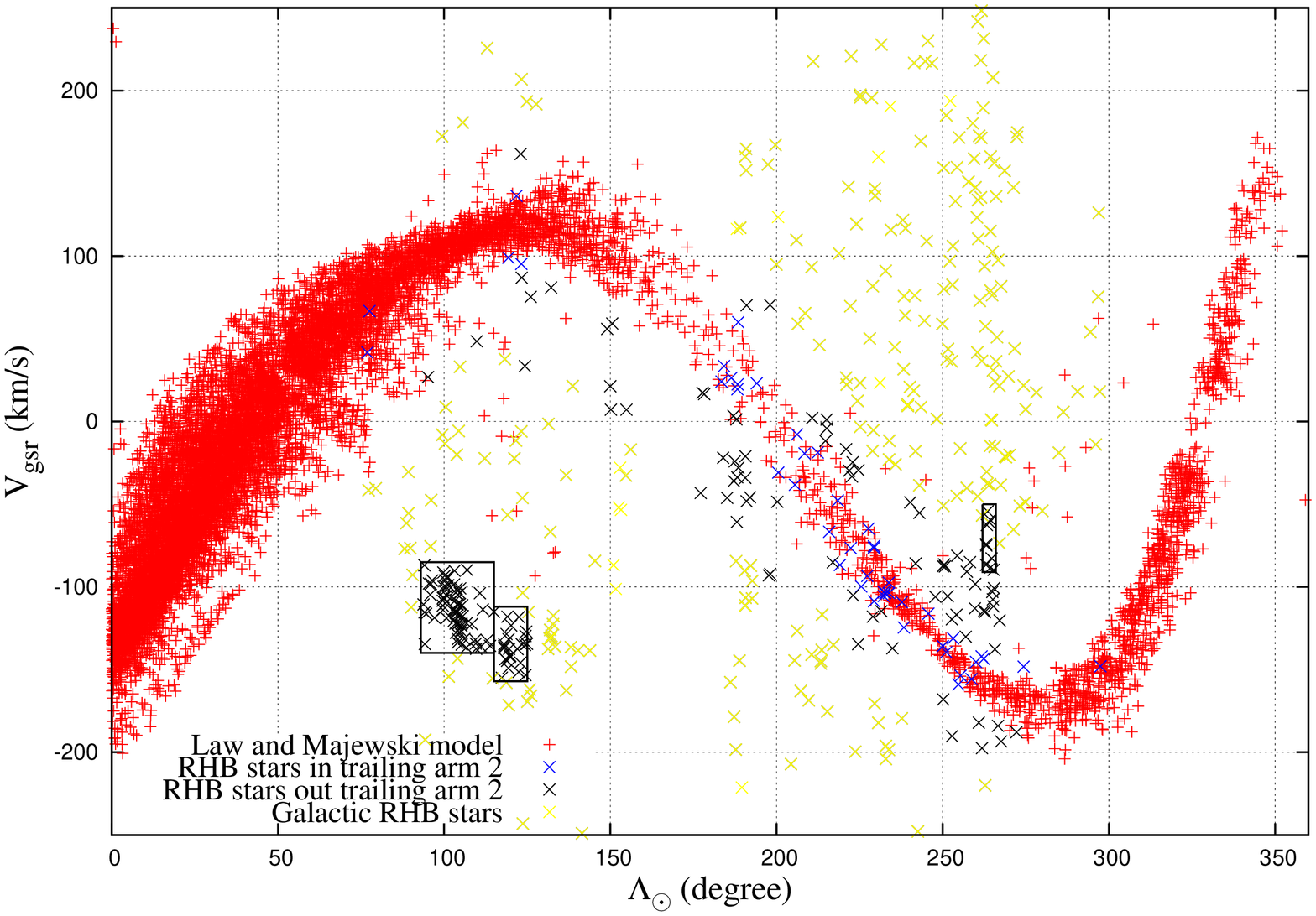}
\end{minipage}
\caption{Selected RHB stars with the \citet{law10} model (red points)
  in a $V_{gsr}$-$\Lambda_\odot$ map. We obtain 118 stars (green
  points) from 586 stars (green points + yellow points + black points)
  in leading arm 1 and 80 stars (green points) from 585 stars (green
  points + yellow points + black points) in leading arm 2. We obtain
  329 stars (blue points) from 973 stars (blue points + yellow points
  + black points) in trailing arm 1 and 47 stars (blue points) from
  502 stars (blue points + yellow points + black points) in trailing
  arm 2. The yellow points indicate the Galactic RHB stars. We
  replaced the high density stars for a generalization which is shown
  as the black boxes. For clearly dividing the stars into Sgr and
  Galaxy components, we omit the stars (black points) located at the edge 
  of the model.}
\label{fig:vgsrLambda}
\end{figure*}

\begin{figure*}[!htbp]
\begin{minipage}[t]{0.5\textwidth}
\centering
{\bf Leading arm stars in Sgr}
\end{minipage}
\begin{minipage}[t]{0.5\textwidth}
\centering
{\bf Trailing arm stars in Sgr}
\end{minipage}
\begin{minipage}{0.5\textwidth}
\centering
\includegraphics[angle=-90,width=80mm]{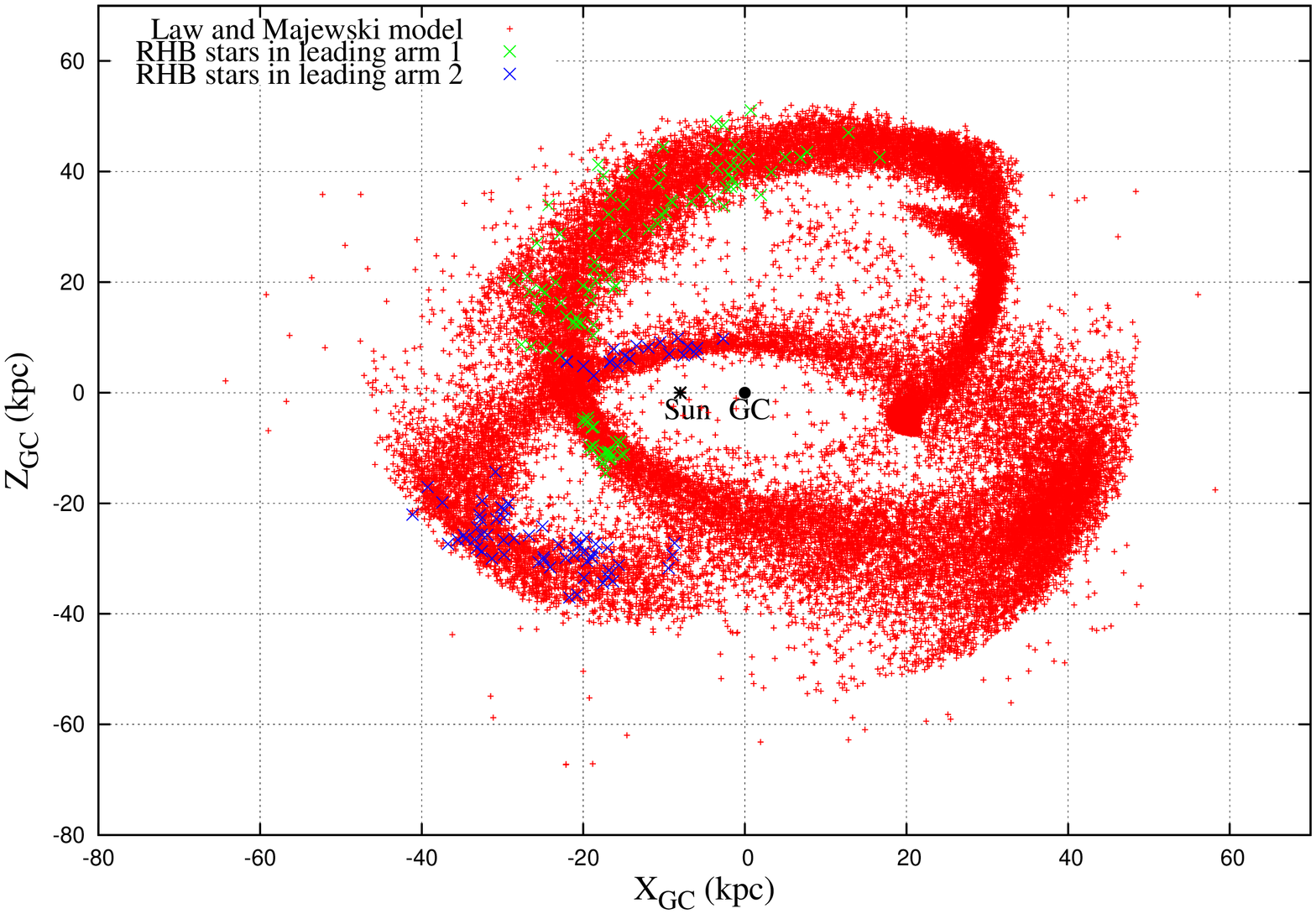}
\end{minipage}
\begin{minipage}{0.5\textwidth}
\centering
\includegraphics[angle=-90,width=80mm]{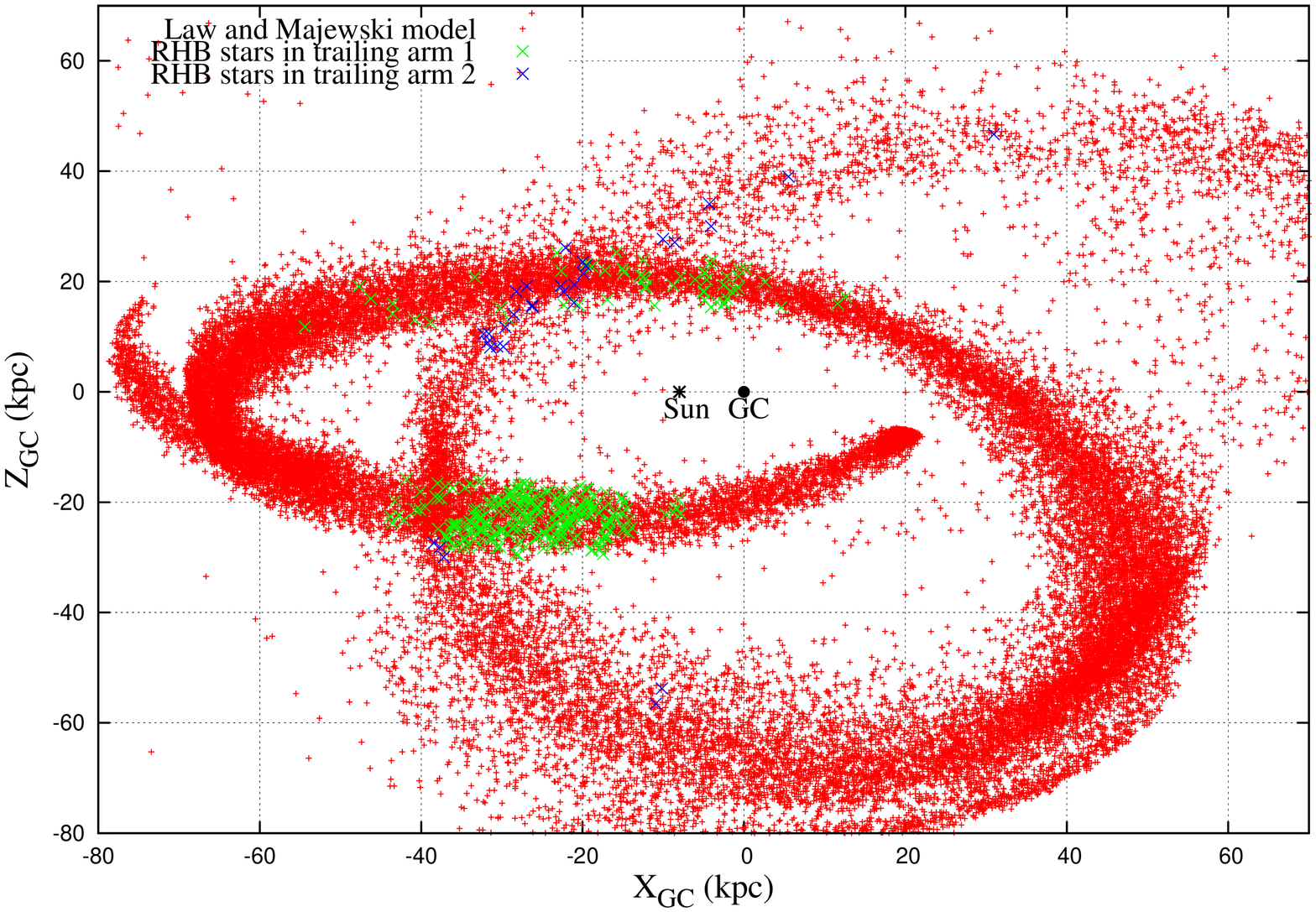}
\end{minipage}\vspace{0.2cm}
\begin{minipage}[t]{0.5\textwidth}
\centering
{\bf Leading arm stars in Galaxy}
\end{minipage}
\begin{minipage}[t]{0.5\textwidth}
\centering
{\bf Trailing arm stars in Galaxy}
\end{minipage}
\begin{minipage}{0.5\textwidth}
\centering
\includegraphics[angle=-90,width=80mm]{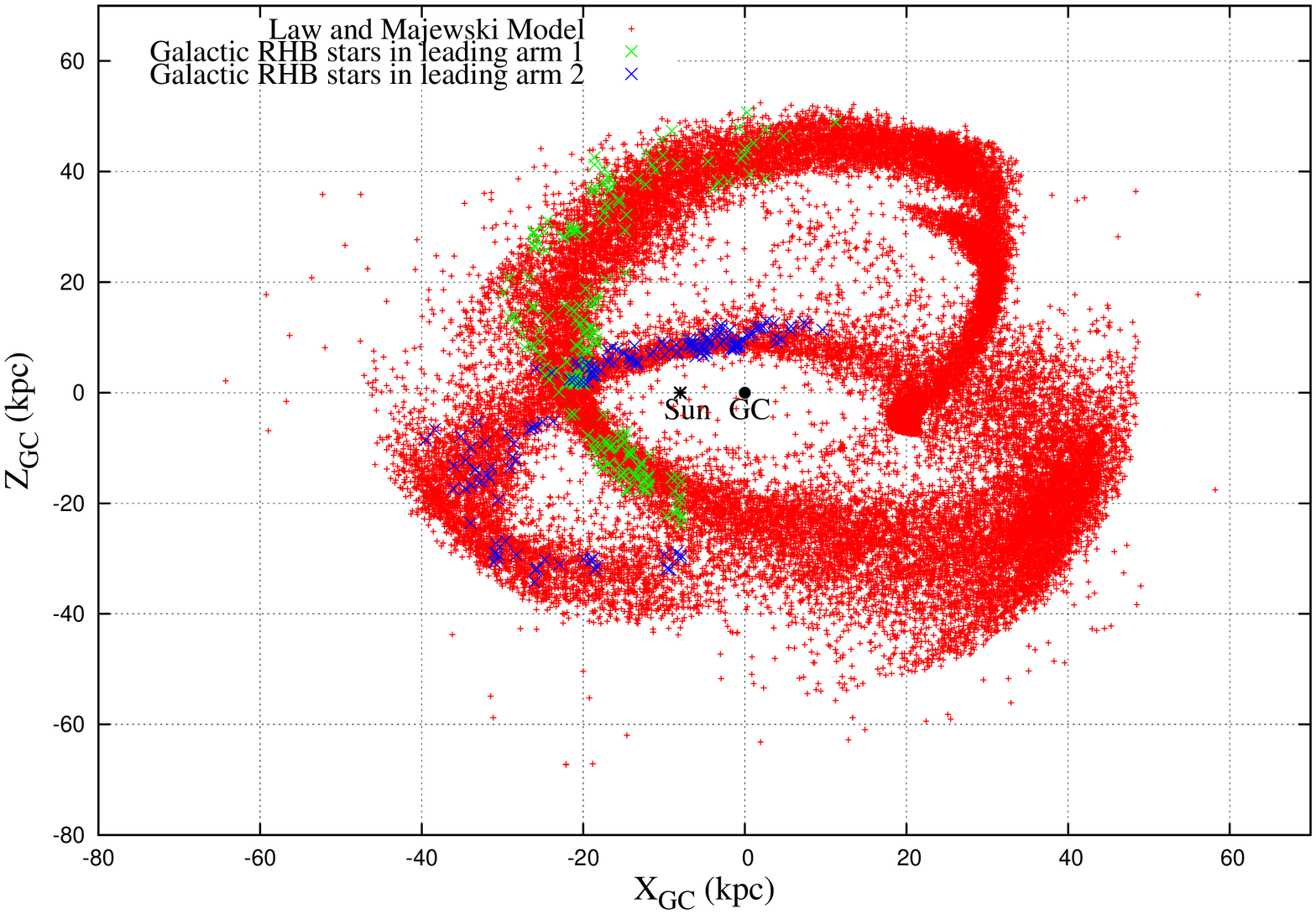}
\end{minipage}
\begin{minipage}{0.5\textwidth}
\centering
\includegraphics[angle=-90,width=80mm]{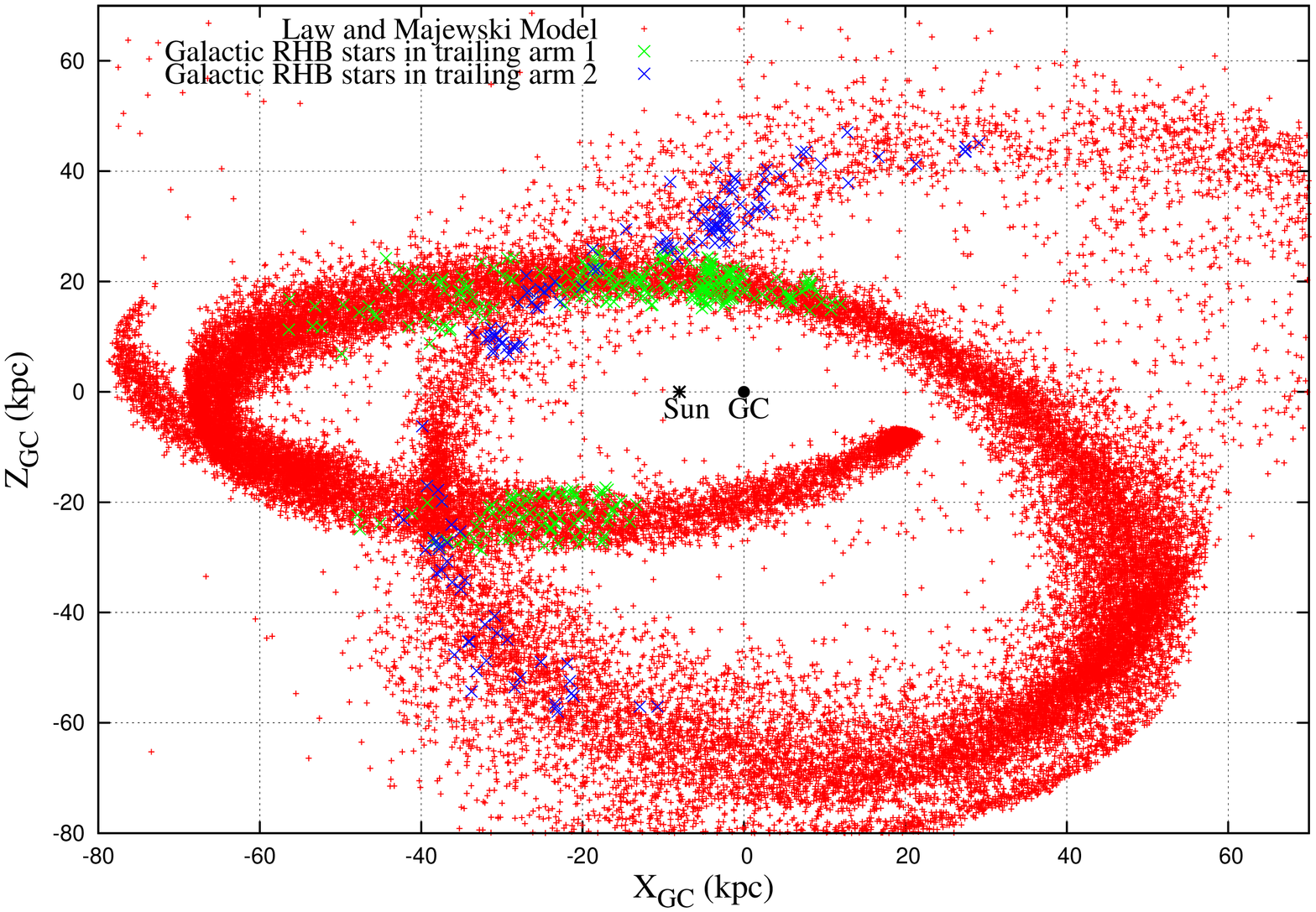}
\end{minipage}
\caption{Spatial distribution of target stars in the debris streams of
  Sgr. Upper panels: A plot of Sgr RHB stars in the leading and
  trailing arms in an $X_{GC}$-$Z_{GC}$ map. Green and blue points
  indicate arm 1 and arm 2 respectively in the leading and trailing
  arms. The large black point indicates the location of the Galactic
  Center, while the asterisk indicates the location of the Sun. Lower
  panels: A plot of the Galactic RHB stars in an $X_{GC}$-$Z_{GC}$
  map.}
\label{fig:xz}
\end{figure*}

We select our sample stars by using the \citet{law10} model as a
reference to provide cuts on the RHB stars. Firstly, we obtain 3512
stars from the full 8535 RHB star sample using Ra-Dec positions.
Secondly, we choose RHB stars in the Sagittarius leading and trailing
tidal tails using a Distance-$\Lambda_\odot$ map of the \citet{law10}
model. Here $\Lambda_\odot$ is the Sgr longitude scale along the
orbital plane. We obtain 586 stars in leading arm 1, 585 stars in
leading arm 2, 973 stars in trailing arm 1 and 502 stars in trailing
arm 2 from the 3512 stars. The first and second wrap of the
\citet{law10} model is denoted by arm 1 and 2 respectively. Third,
we select stars to be likely members of the Sgr stream based
on their radial velocities, which are appropriate for the Sgr stream at these
positions based on Sgr debris models \citep{cho10}. Specifically, we
select stars with a $V_{gsr}$(the velocity in the Galactic standard of
rest)-$\Lambda_\odot$ map (Figure~\ref{fig:vgsrLambda}). A local
standard of rest rotation velocity of 220 km~s$^{-1}$ is adopted for
the Sun, for consistency with \citet{law10}. We also calculate
$V_{gsr}$ with the same equation as \citet{law10} for consistency,
i.e. $V_{gsr}=rv + 9.0 \cos b \cos l + 232.0 \cos b \sin l + 7.0 \sin
b$ km~s$^{-1}$. With the $V_{gsr}$ criteria, 118 stars satisfy the
cuts on leading arm 1 in both the Distance-$\Lambda_\odot$ and
$V_{gsr}$-$\Lambda_\odot$ maps from 586 stars. In a similar way, 80,
329 and 47 stars in the leading arm group 2, trailing arm group 1 and
trailing arm group 2, are selected from 585, 973 and 502 stars,
respectively (see Figure~\ref{fig:vgsrLambda}). There are 18 RHB stars
overlapped in the leading arm and trailing arm. We omit these
overlapping stars from the leading and trailing arm groups. Finally
there are 102, 78, 327 and 31 RHB stars in leading arm 1 and 2 and
trailing arm 1 and 2, respectively (shown in the $X_{GC}$-$Z_{GC}$ map of
Figure~\ref{fig:xz}). We adopt 556 (including 18 overlapping stars) as the
total number of RHB stars in our Sgr samples. The table of our Sgr 
samples is provided in a electronic version.

For comparison, we selected a sample of Galactic stars with the same
positions but different velocities from the Sgr tidal stars. That is,
we select Galactic RHB stars from the full 8535 star sample by finding
stars that satisfy the Ra-Dec criteria and the
Distance-$\Lambda_\odot$ criteria of the \citet{law10} model, but do
not satisfy the $V_{gsr}$-$\Lambda_\odot$ criteria mentioned above. In
Figure~\ref{fig:vgsrLambda} the yellow points indicate Galactic RHB
stars. We excluded the high density stars with a generalization which
is shown as the black boxes in order to reduce the potential effect
from the undetected stream stars in the Galaxy, see
Figure~\ref{fig:vgsrLambda}. Again, 159 overlapping stars at the
positions of the leading and trailing arms are removed from the
leading and trailing sample. Finally, there are 202, 164, 347 and 129
Galactic stars at the positions of leading arm 1 and 2 and trailing
arm 1 and 2, respectively (shown in the $X_{GC}$-$Z_{GC}$ map of
Figure~\ref{fig:xz}). We adopt 1001 (including 159 overlapping stars)
as the total number of RHB stars in our Galaxy sample. The table of Galactic
samples is also provided in a electronic version.

\section{Results and Discussions}
\subsection{Comparing the kinematics and chemistry of RHB stars in the
  Sgr and in the Galaxy}
\begin{figure*}[!htbp]
\begin{minipage}[t]{0.5\textwidth}
\centering
{\bf All RHB stars in Sgr}
\end{minipage}
\begin{minipage}[t]{0.5\textwidth}
\centering
{\bf All RHB stars in Galaxy}
\end{minipage}
\begin{minipage}{\textwidth}
\includegraphics[angle=-90,width=75mm]{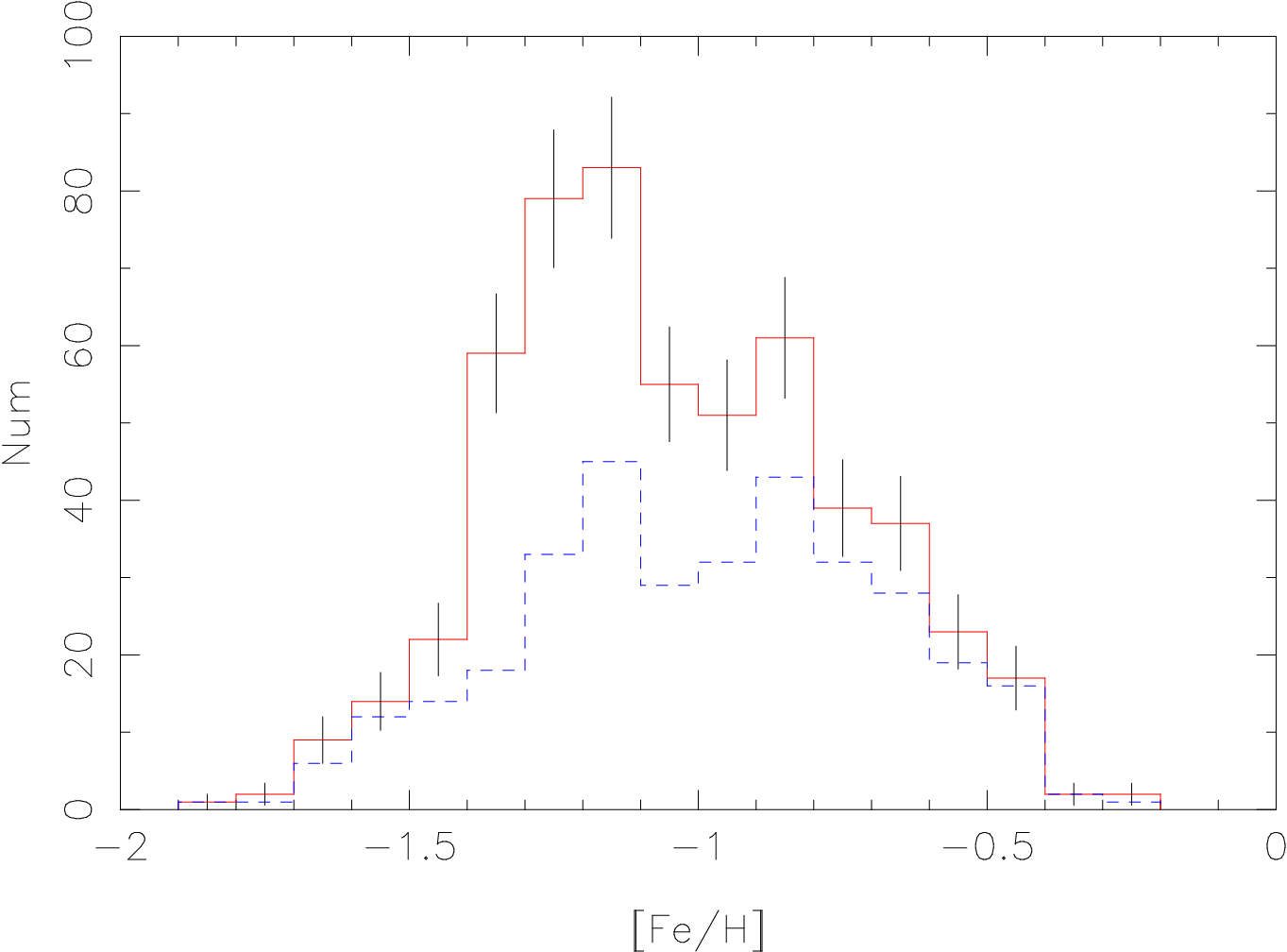}
\hspace*{\fill}
\includegraphics[angle=-90,width=75mm]{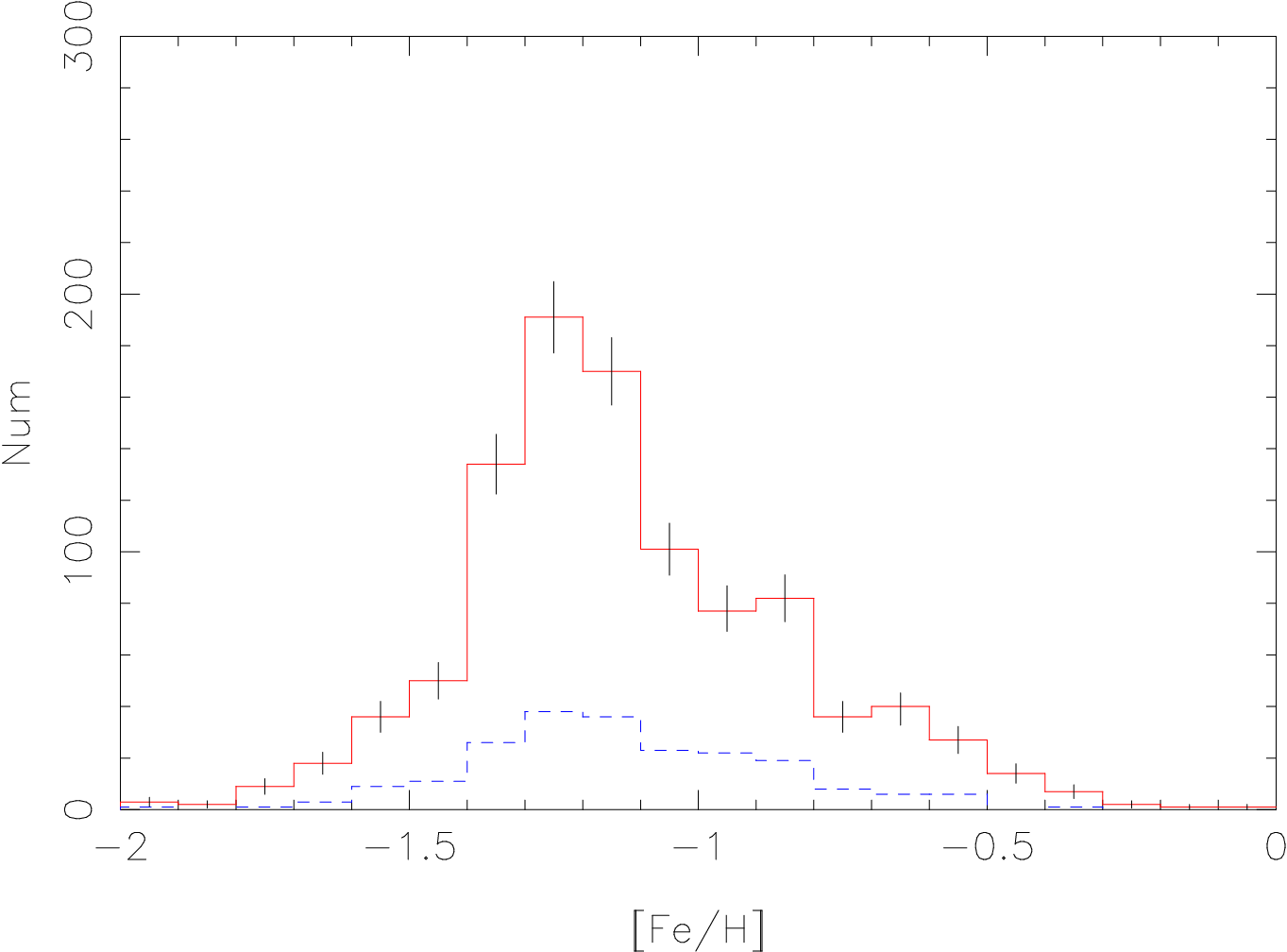}
\end{minipage}\vspace{0.2cm}
\begin{minipage}[t]{0.5\textwidth}
\centering
{\bf }
\end{minipage}
\begin{minipage}[t]{0.5\textwidth}
\centering
{\bf }
\end{minipage}
\begin{minipage}{\textwidth}
\includegraphics[angle=-90,width=75mm]{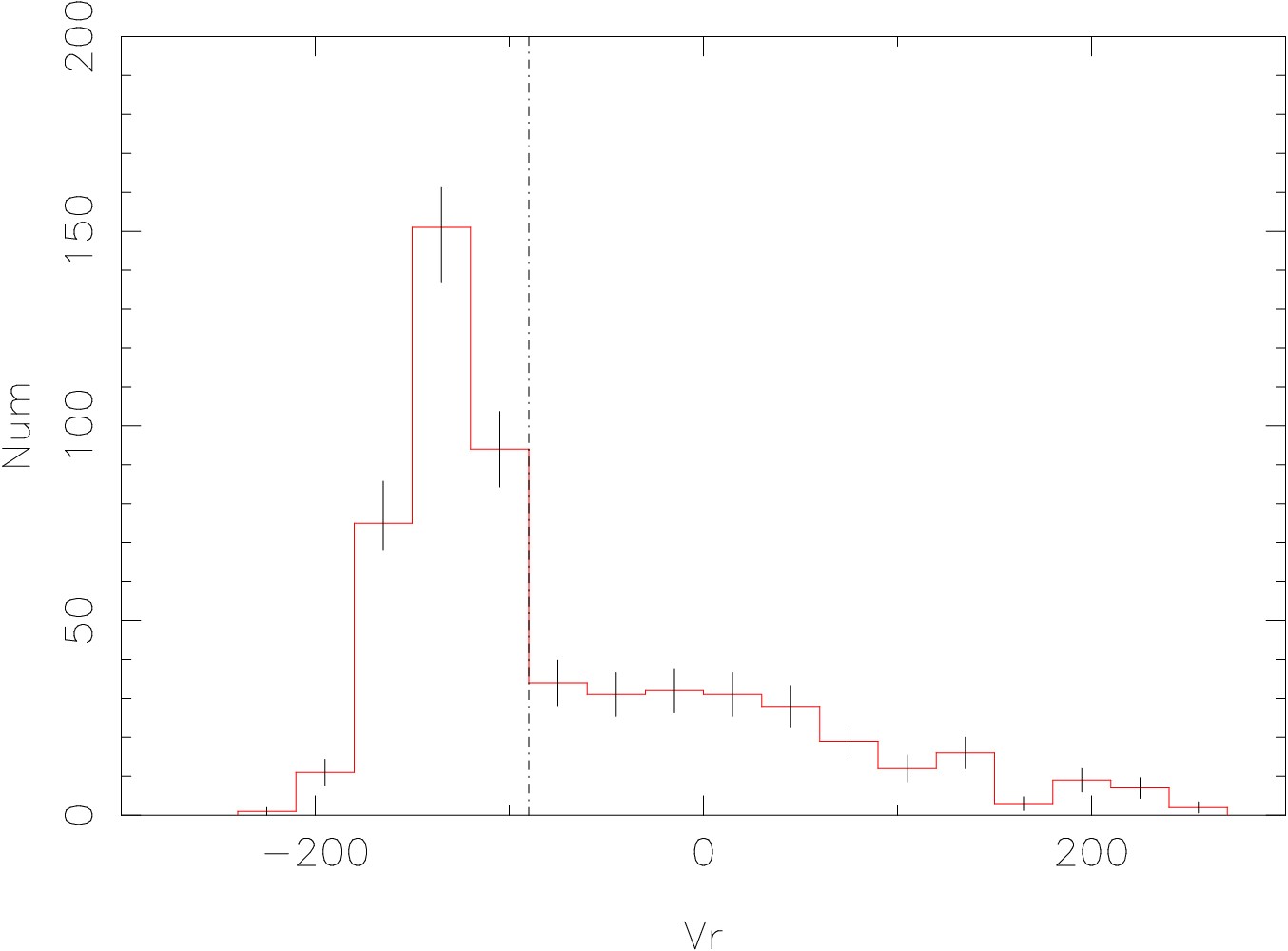}
\hspace*{\fill}
\includegraphics[angle=-90,width=75mm]{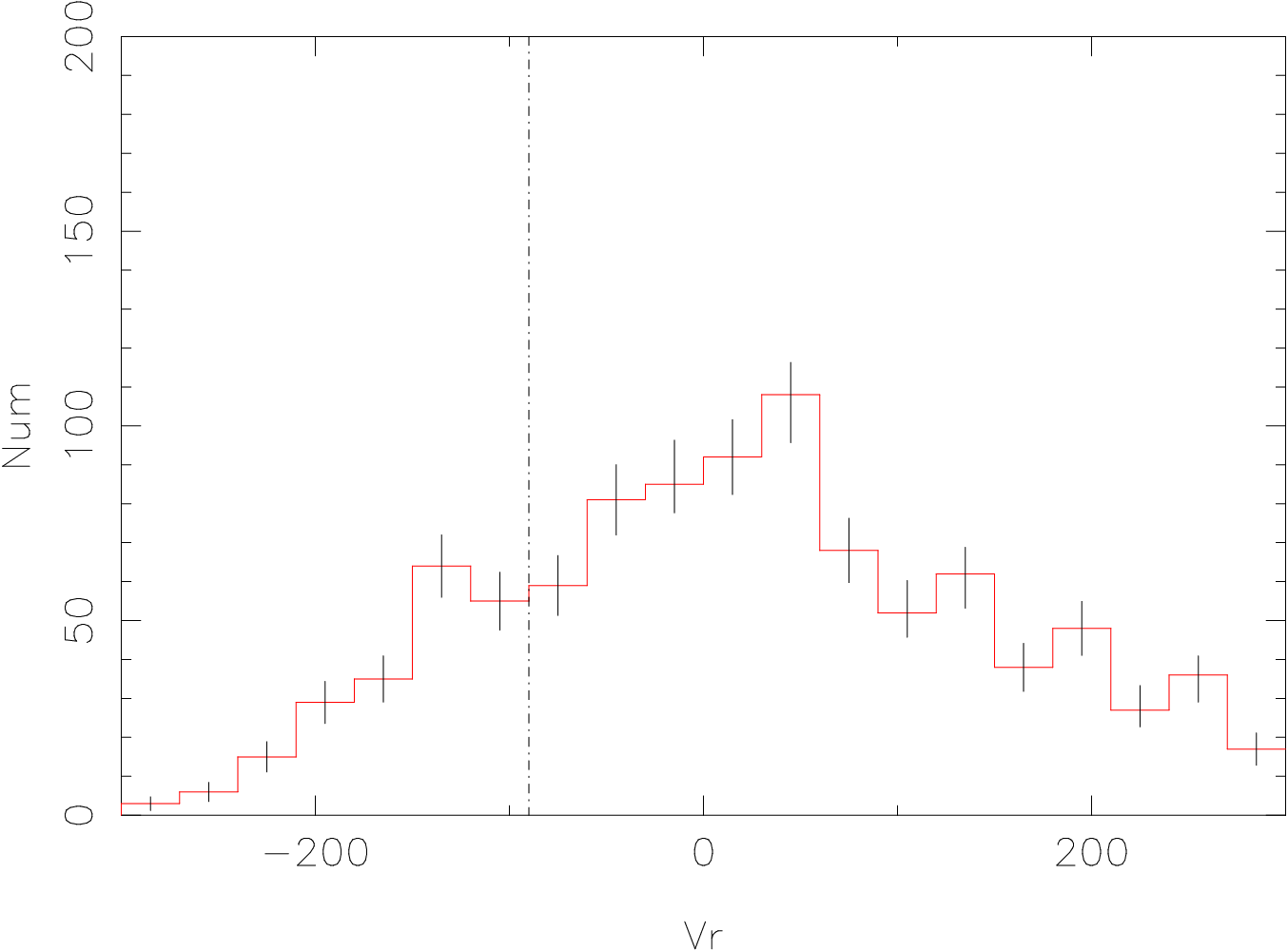}
\end{minipage}\vspace{0.2cm}
\begin{minipage}[t]{0.5\textwidth}
\centering
{\bf }
\end{minipage}
\begin{minipage}[t]{0.5\textwidth}
\centering
{\bf }
\end{minipage}
\begin{minipage}{\textwidth}
\includegraphics[angle=-90,width=75mm]{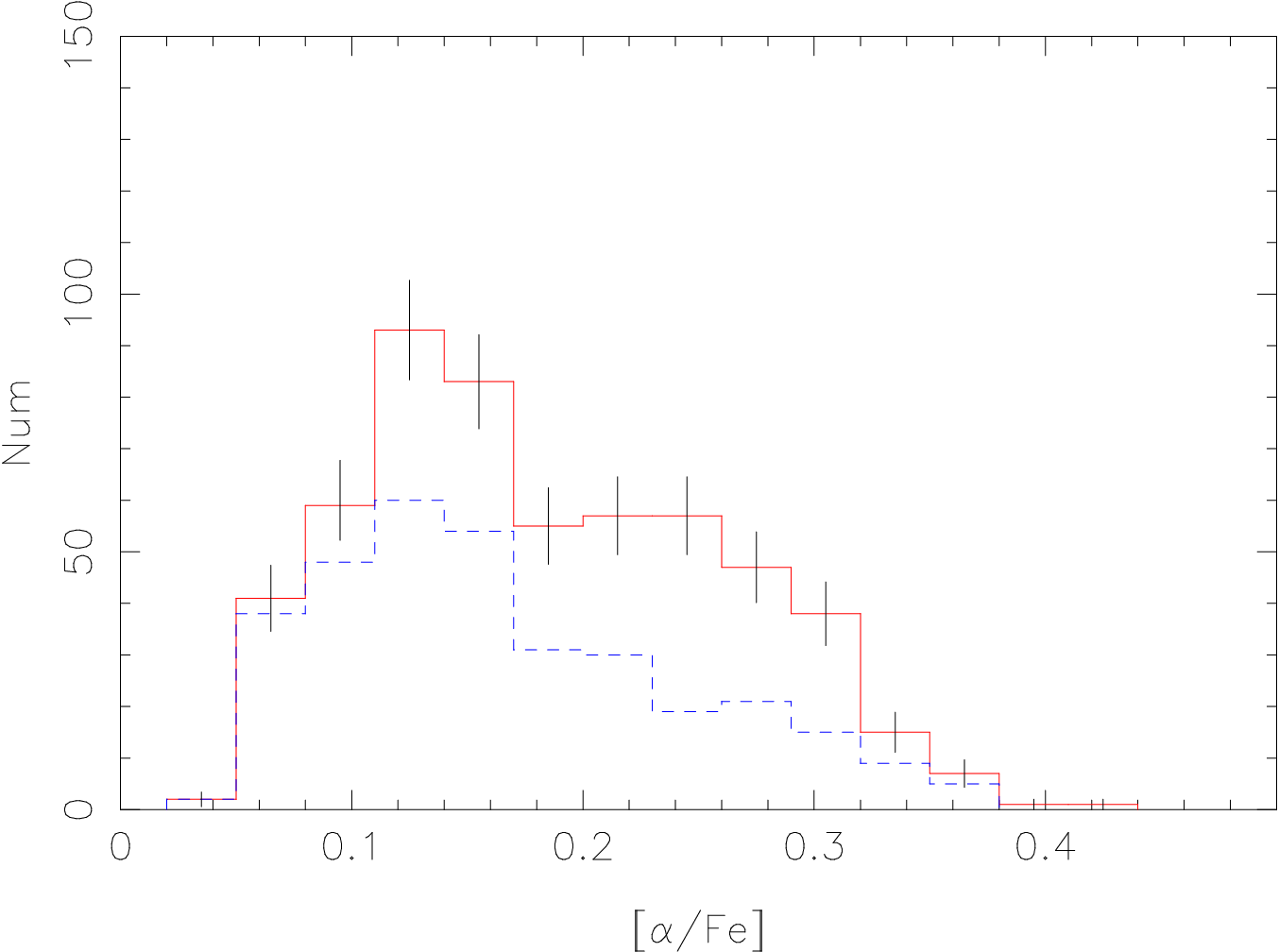}
\hspace*{\fill}
\includegraphics[angle=-90,width=75mm]{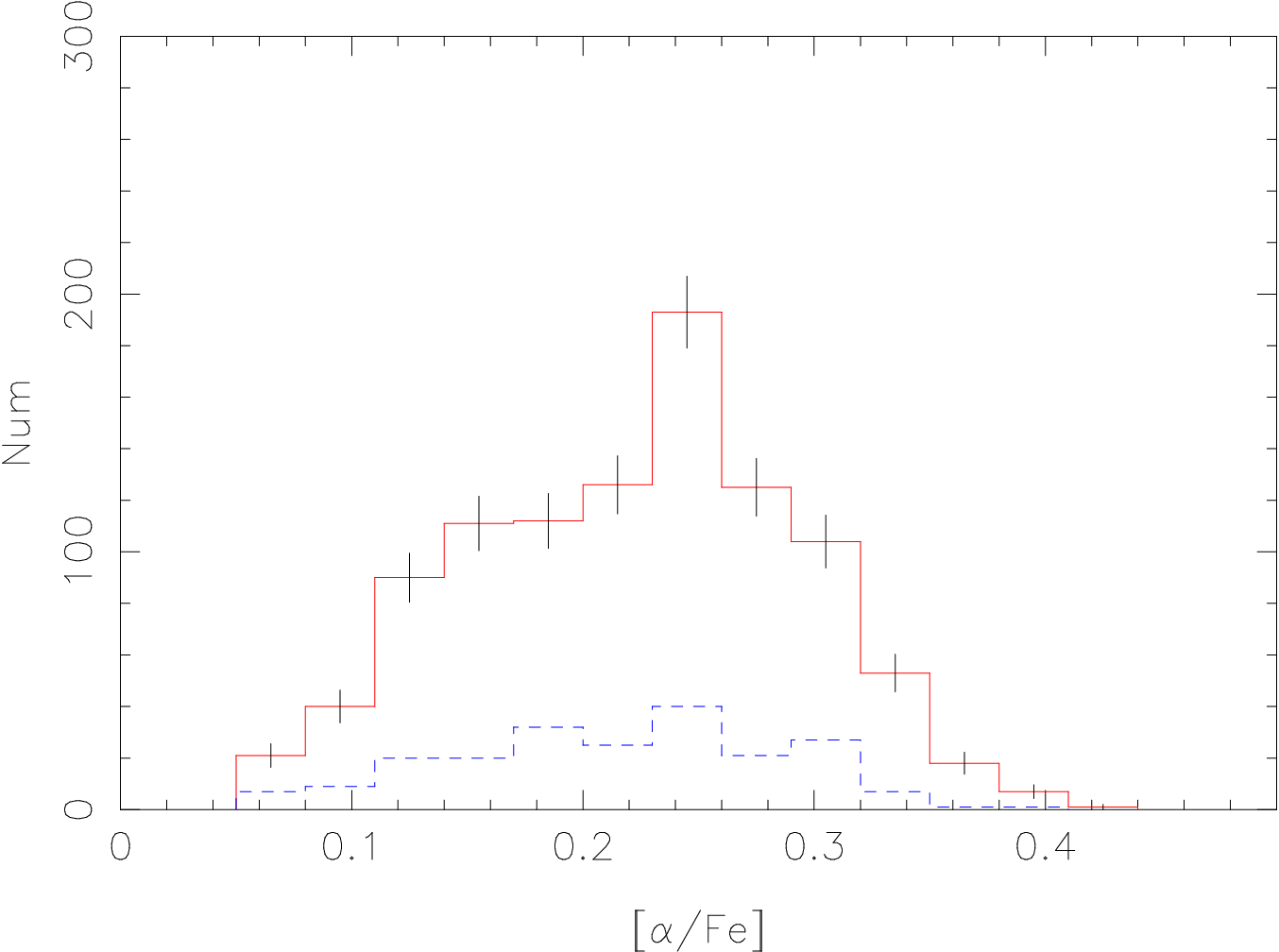}
\end{minipage}\vspace{0.2cm}
\begin{minipage}[t]{0.5\textwidth}
\centering
{\bf }
\end{minipage}
\begin{minipage}[t]{0.5\textwidth}
\centering
{\bf }
\end{minipage}
\begin{minipage}{\textwidth}
\includegraphics[angle=-90,width=75mm]{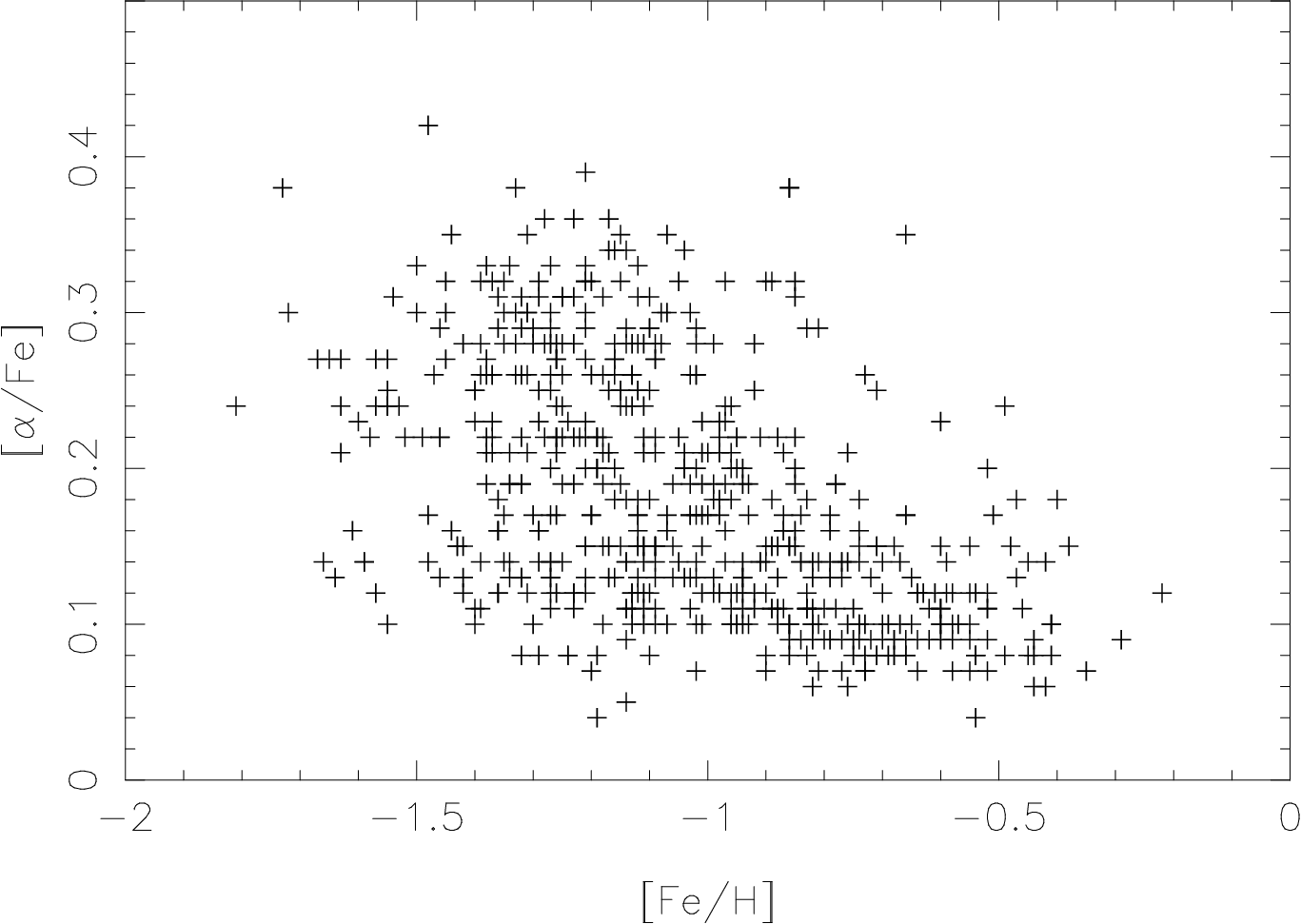}
\hspace*{\fill}
\includegraphics[angle=-90,width=75mm]{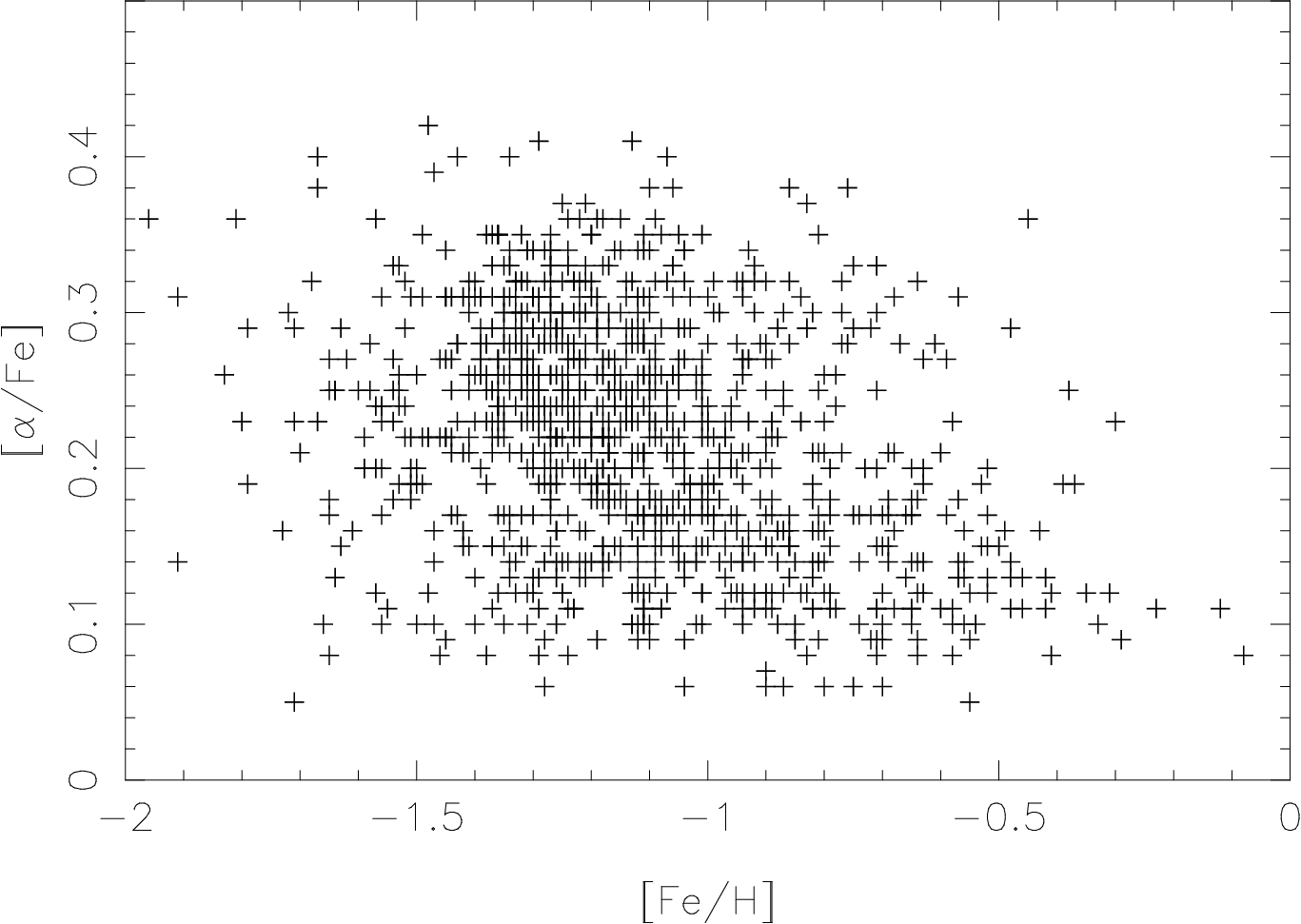}
\end{minipage}
\caption{We compare all RHB stars (red solid lines) in the Sgr (556)
  and Galaxy (1001). Blue dashed lines show the stars whose $V_r$ is
  less than -90\,km~s$^{-1}$. Left (right) panels show RHB stars in the
  Sgr (Galaxy).}
\label{fig:allHisto}
\end{figure*}

In Figure~\ref{fig:allHisto}, we plot the histograms of [Fe/H], $V_r$,
and [$\alpha$/Fe] and the distribution map of [Fe/H]-[$\alpha$/Fe] for
RHB stars in the Sgr (556) and Galaxy (1001). One sees that the value
of $V_r$ for all RHB stars in the Sgr have a sharp peak at
-140~km~s$^{-1}$, while there is a large dispersion in the
distribution for the Galactic RHB stars. This is mainly due to the
selection effect. We set a dashed line at -90~km~s$^{-1}$ in the $V_r$
histogram to define the lower velocity group and analyze the metallity
distribution of those lower velocity stars. Clearly, the lower
velocity stars are more than half of all the stars in the Sgr, but the
lower velocity stars are only a small part of all the Galactic
stars. Meanwhile, the Galactic stars are dominated by a metal poor
component while the Sgr stars have a significant contribution from a
more metal rich component. The distribution of [Fe/H] in the Sgr stars
has two peaks, one at -1.3\,dex and one at -0.8\,dex. There are also 
two peaks in the [Fe/H] distribution of Galactic RHB stars, which are at
the same [Fe/H] value, but the peak at -0.8\,dex is less pronounced 
and could be due to Sgr stars near the edges of our selection criteria.  
\citet{yan09} show that the giant branch in the Sgr
leading tidal tail is consistent with those of globular clusters with
[Fe/H] of -1.0~$\pm$~0.5. They also find that the 33 identified Sgr
K/M-giant stars have metallicities of -0.8~$\pm$~0.2. Our results are
similar to the distribution of [Fe/H] in \citet{yan09}. We also show
the lower velocity stars with a dashed line in the histograms of
[Fe/H] and [$\alpha$/Fe]. For the Sgr stars, the dashed line shows two
peaks in the [Fe/H] histogram and the two components have equal
contributions, while the solid line shows a bigger contribution from
the peak at -1.3\,dex than that from the peak at -0.8\,dex. For
Galactic RHB stars, the dashed line is similar to the solid line.

From the [Fe/H]-[$\alpha$/Fe] map in Figure~\ref{fig:allHisto}, we can
see that the [$\alpha$/Fe] of most stars is lower than 0.2\,dex in
Sgr, while that of most Galactic stars is larger than 0.2\,dex. The
low [$\alpha$/Fe] stars mainly come from the metal rich component of
the Sgr tidals at -0.8\,dex. These results are consistent with the
results of early dwarf galaxy fragments. [$\alpha$/Fe] deficiencies
were found by \citet{smecker02}, \citet{mcwillim05},
\citet{sbordone07} and \citet{car10} for the more metal-rich stars in
the Sgr (\citet{mcwilliam10}). The existence of some Galactic stars in
our sample may lead to analysis error, but they could also be real
since there are plenty of examples of Galactic halo stars with low
[$\alpha$/Fe] (e.g. \citet{nissen97} and \citet{brown97}).

\subsection{Comparing the properties of Sgr RHB stars in the
  leading and trailing arms}

\begin{figure*}[!htbp]
\begin{minipage}[t]{0.5\textwidth}
\centering
{\bf Leading arm stars in Sgr}
\end{minipage}
\begin{minipage}[t]{0.5\textwidth}
\centering
{\bf Leading arm stars in Galaxy}
\end{minipage}
\begin{minipage}{\textwidth}
\includegraphics[angle=-90,width=75mm]{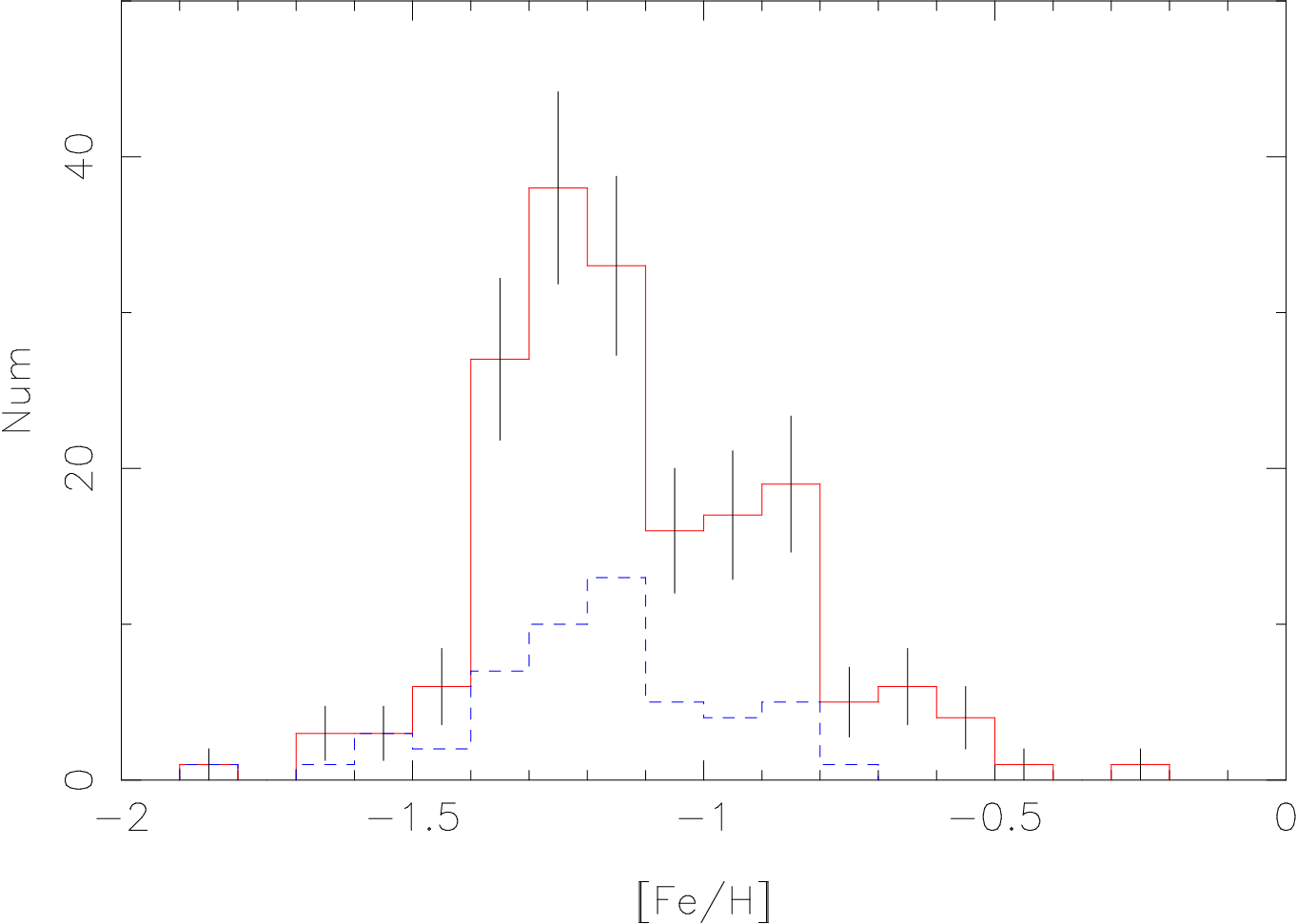}
\hspace*{\fill}
\includegraphics[angle=-90,width=75mm]{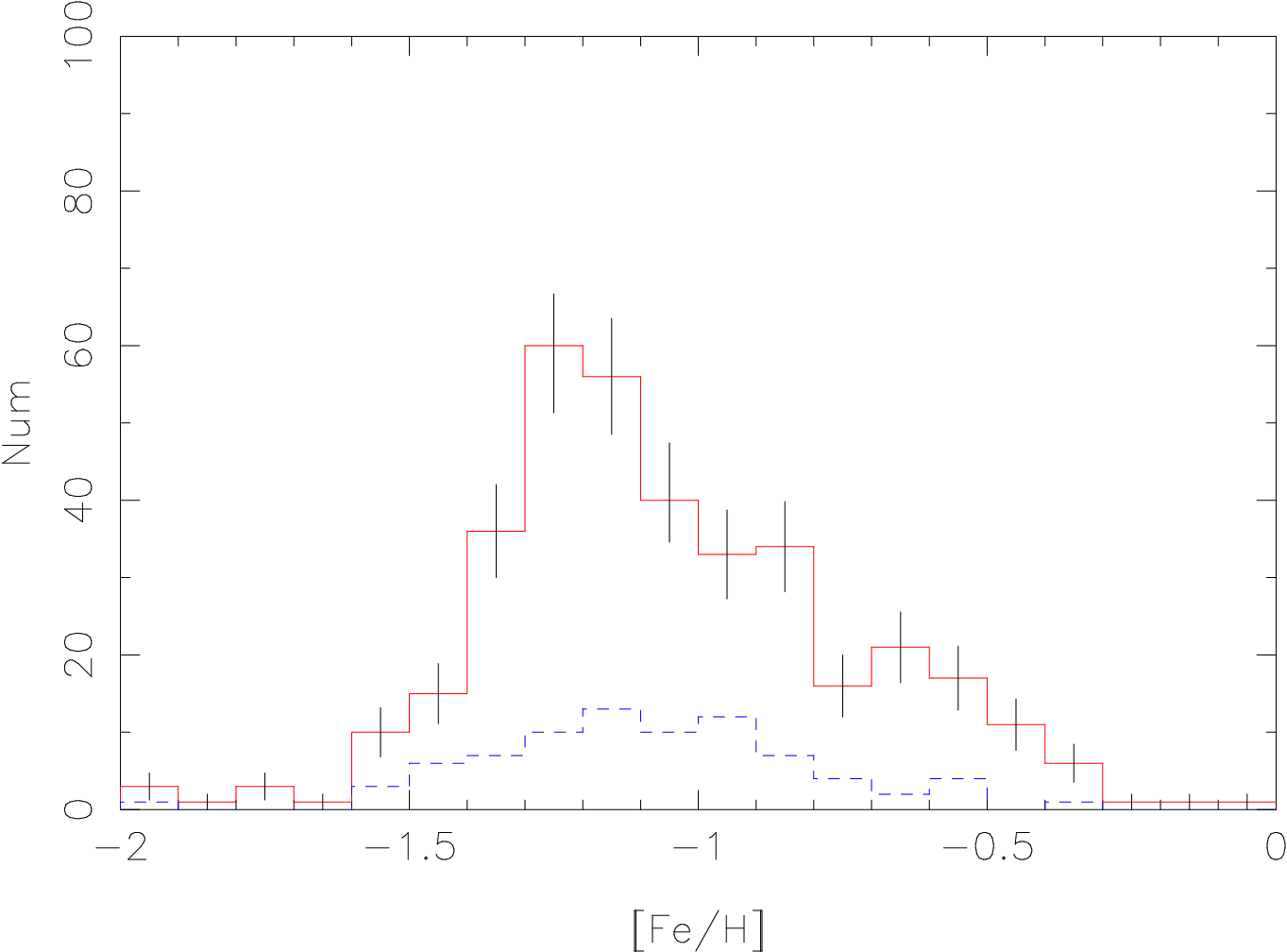}
\end{minipage}\vspace{0.2cm}
\begin{minipage}[t]{0.5\textwidth}
\centering
{\bf }
\end{minipage}
\begin{minipage}[t]{0.5\textwidth}
\centering
{\bf }
\end{minipage}
\begin{minipage}{\textwidth}
\includegraphics[angle=-90,width=75mm]{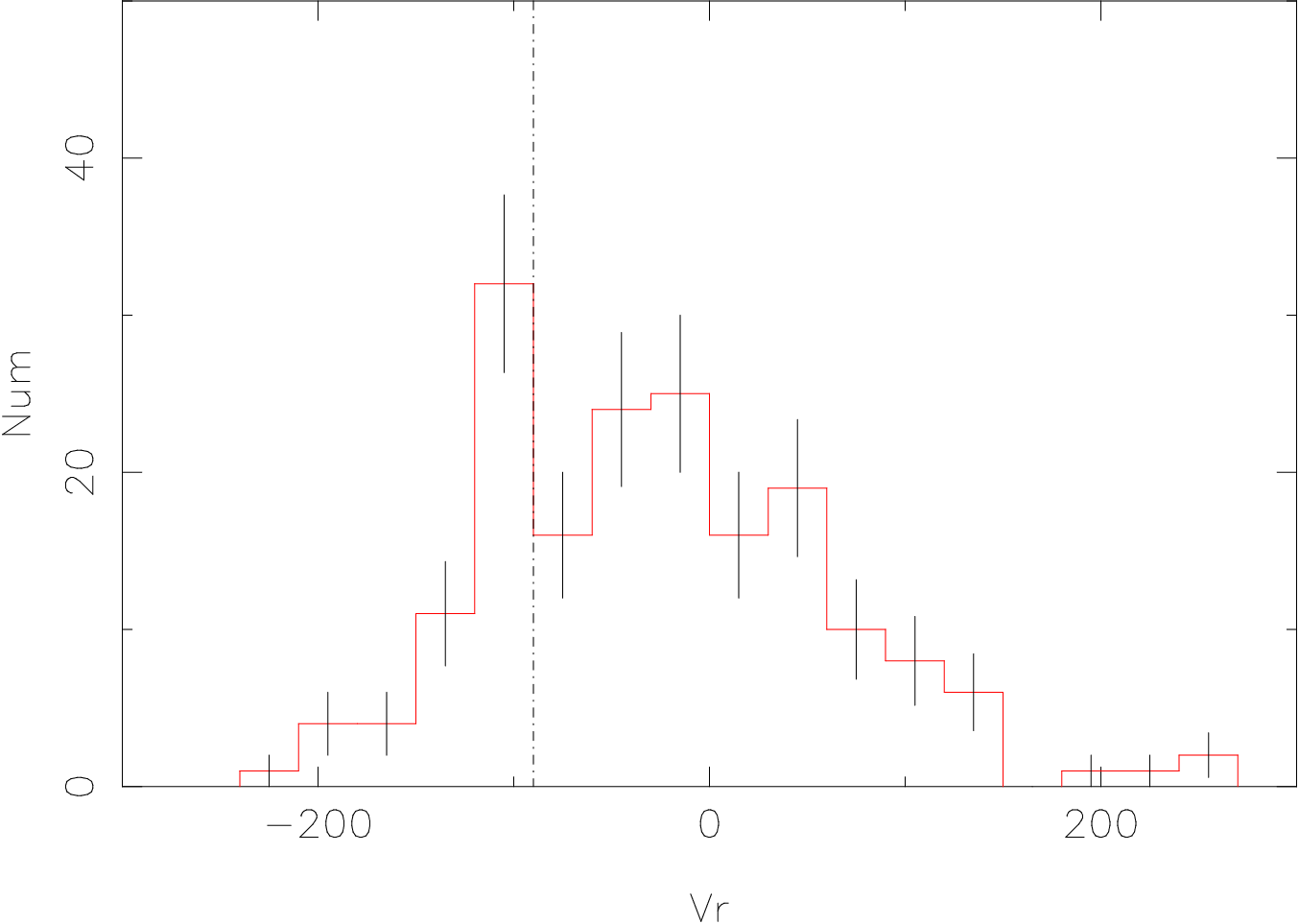}
\hspace*{\fill}
\includegraphics[angle=-90,width=75mm]{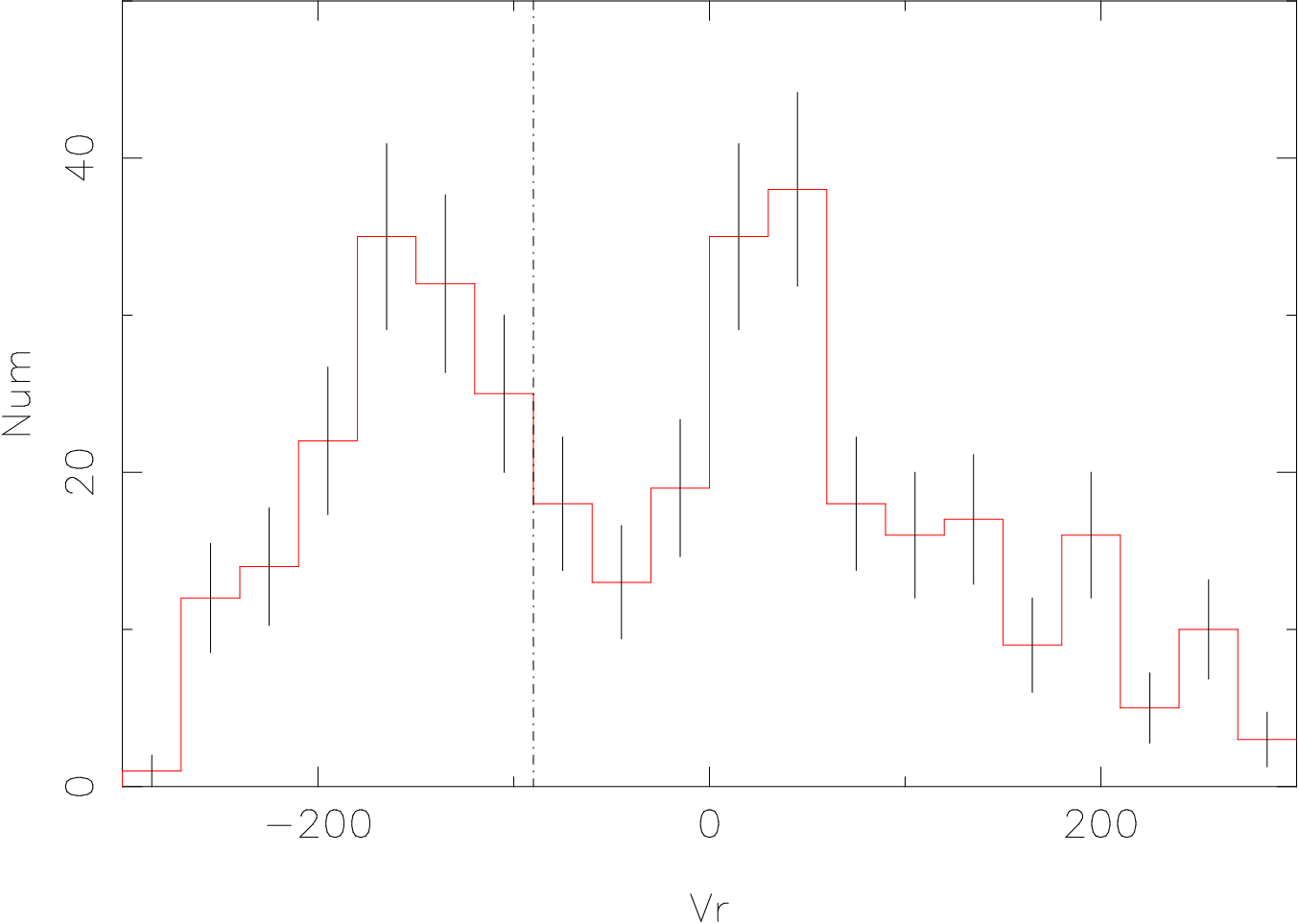}
\end{minipage}\vspace{0.2cm}
\begin{minipage}[t]{0.5\textwidth}
\centering
{\bf }
\end{minipage}
\begin{minipage}[t]{0.5\textwidth}
\centering
{\bf }
\end{minipage}
\begin{minipage}{\textwidth}
\includegraphics[angle=-90,width=75mm]{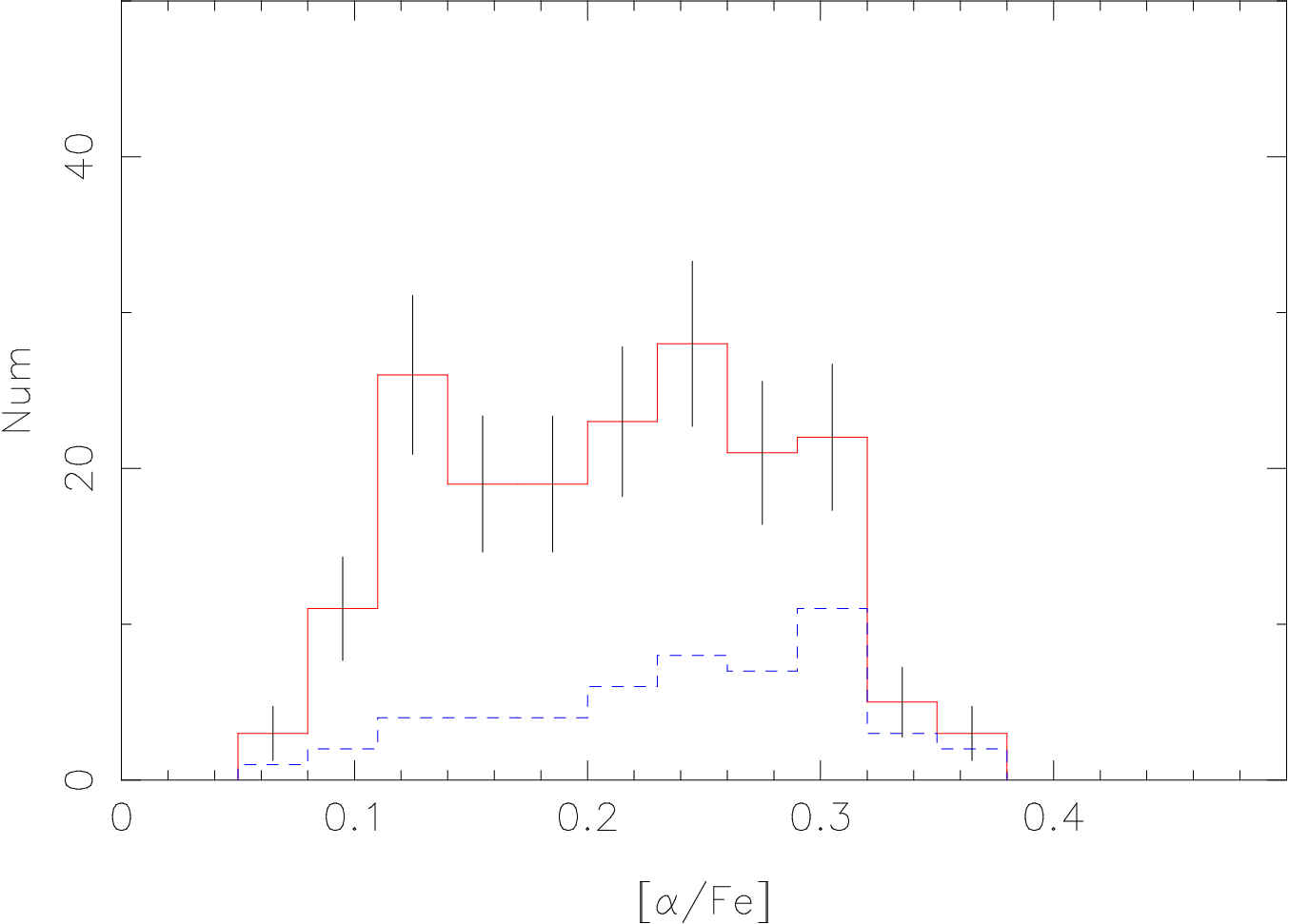}
\hspace*{\fill}
\includegraphics[angle=-90,width=75mm]{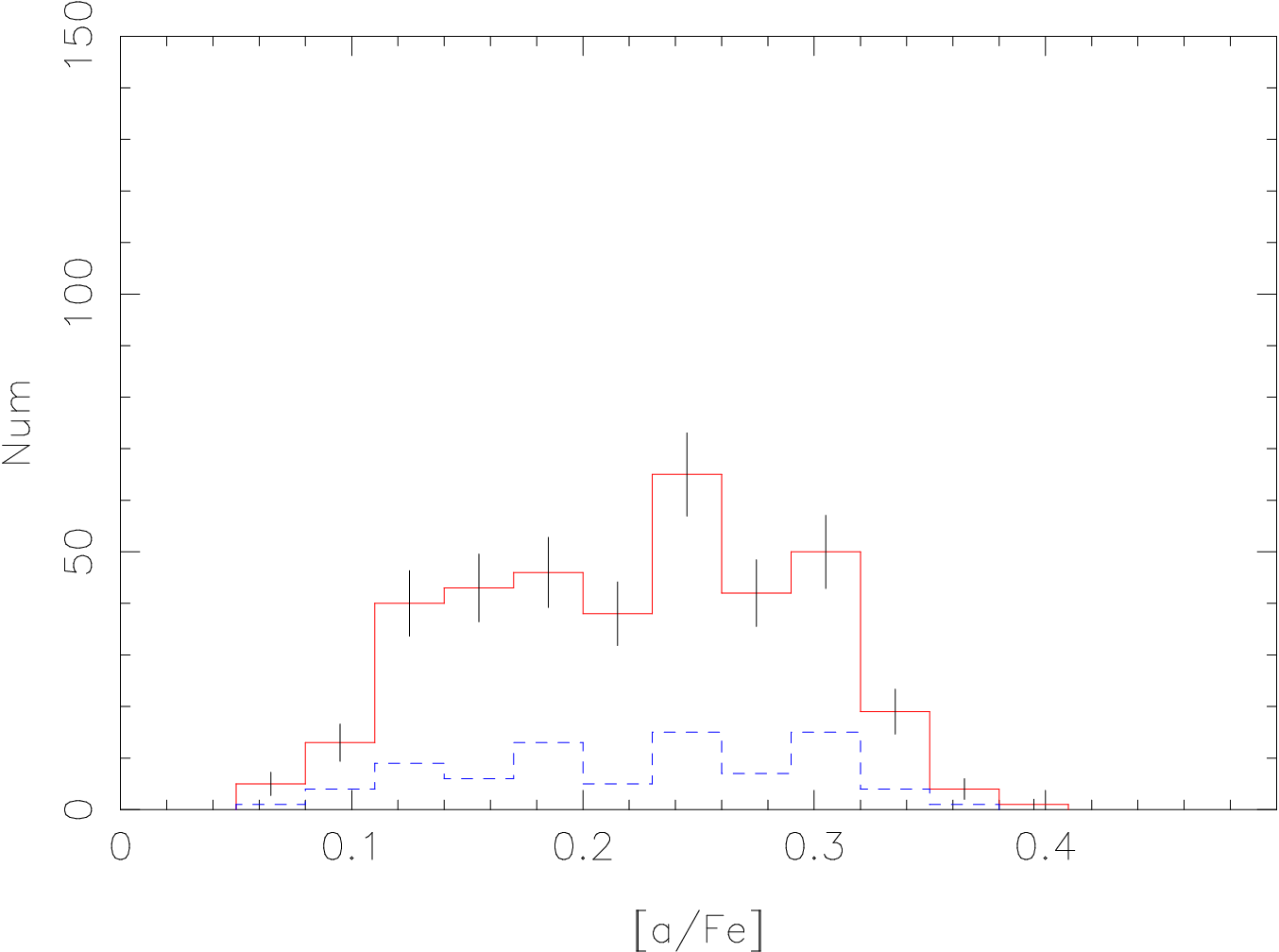}
\end{minipage}\vspace{0.2cm}
\begin{minipage}[t]{0.5\textwidth}
\centering
{\bf }
\end{minipage}
\begin{minipage}[t]{0.5\textwidth}
\centering
{\bf }
\end{minipage}
\begin{minipage}{\textwidth}
\includegraphics[angle=-90,width=75mm]{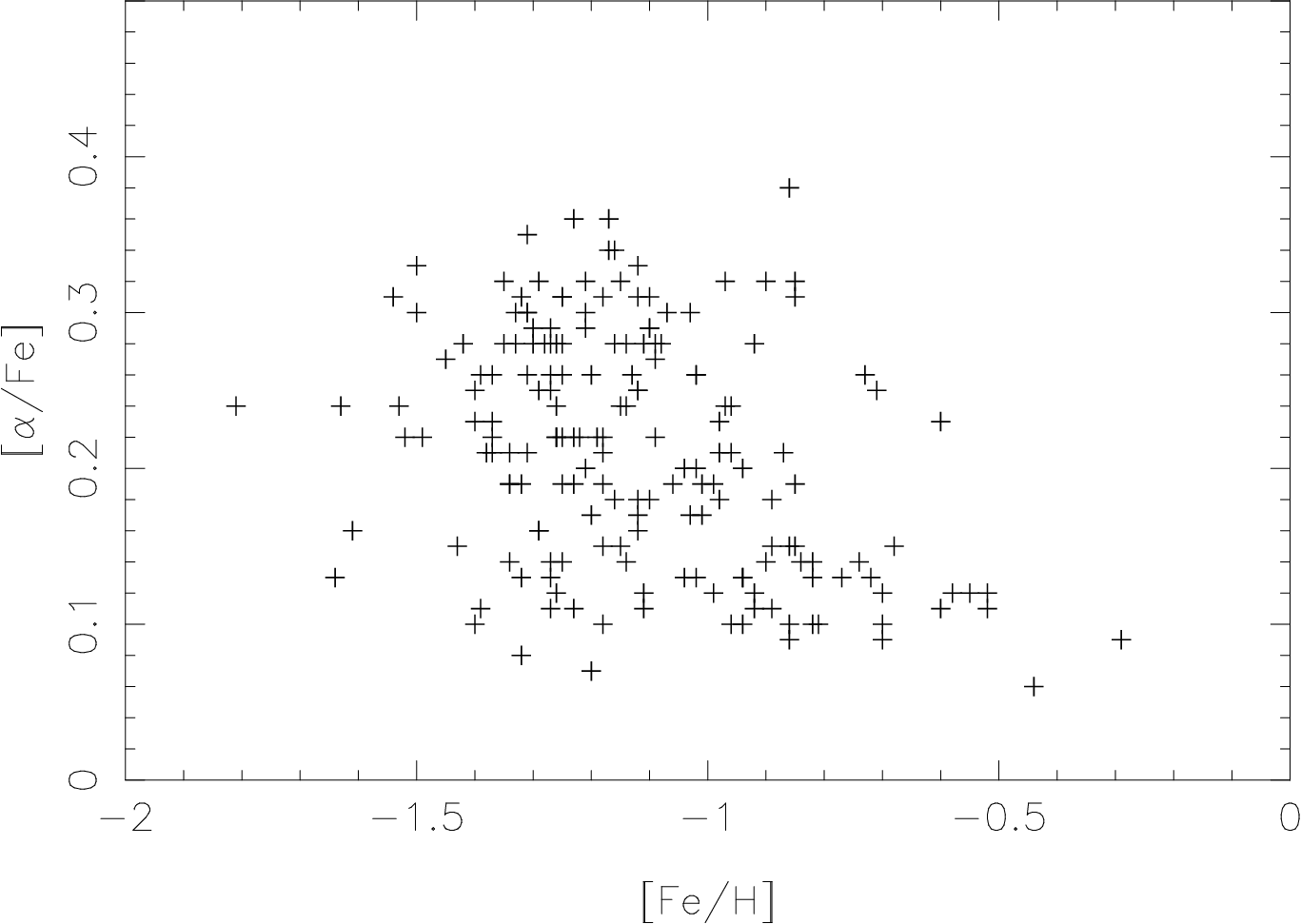}
\hspace*{\fill}
\includegraphics[angle=-90,width=75mm]{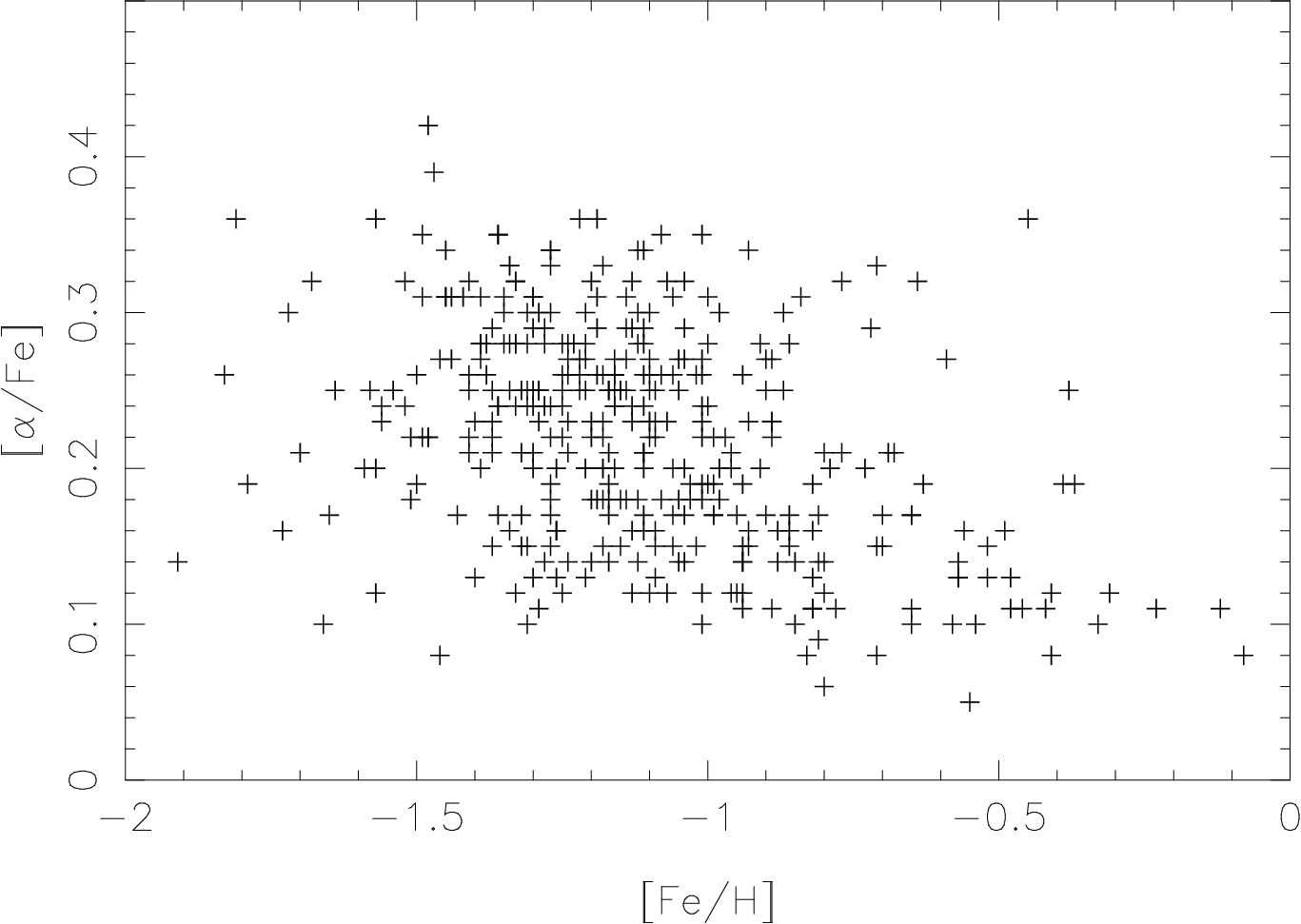}
\end{minipage}
\caption{Same as Figure~\ref{fig:allHisto} but for comparing 
  RHB stars in the leading arm of the Sgr (180) and Galaxy (366).}
\label{fig:leading}
\end{figure*}

\begin{figure*}[!htbp]
\begin{minipage}[t]{0.5\textwidth}
\centering
{\bf Trailing arm stars in Sgr}
\end{minipage}
\begin{minipage}[t]{0.5\textwidth}
\centering
{\bf Trailing arm stars in Galaxy}
\end{minipage}
\begin{minipage}{\textwidth}
\includegraphics[angle=-90,width=75mm]{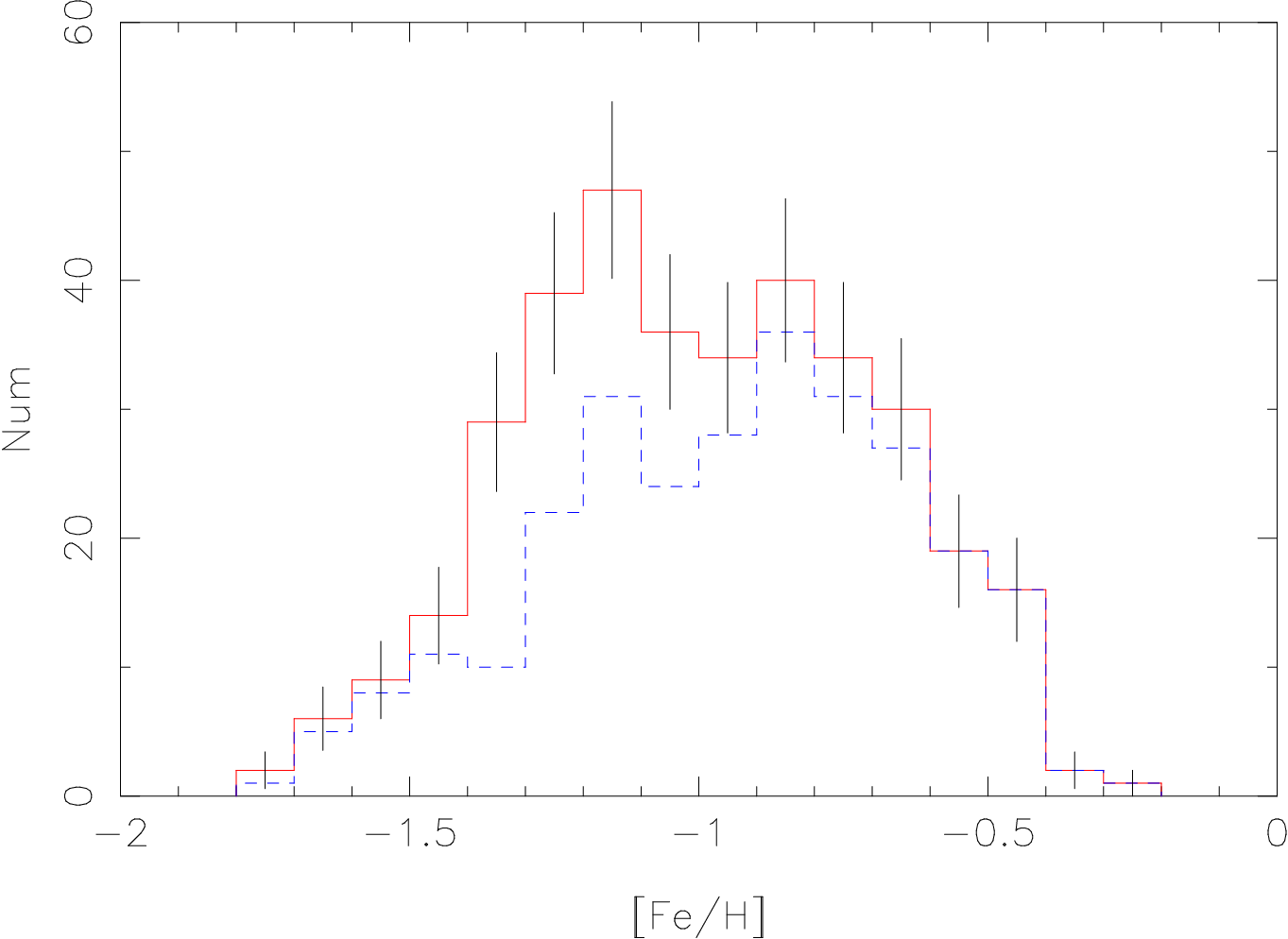}
\hspace*{\fill}
\includegraphics[angle=-90,width=75mm]{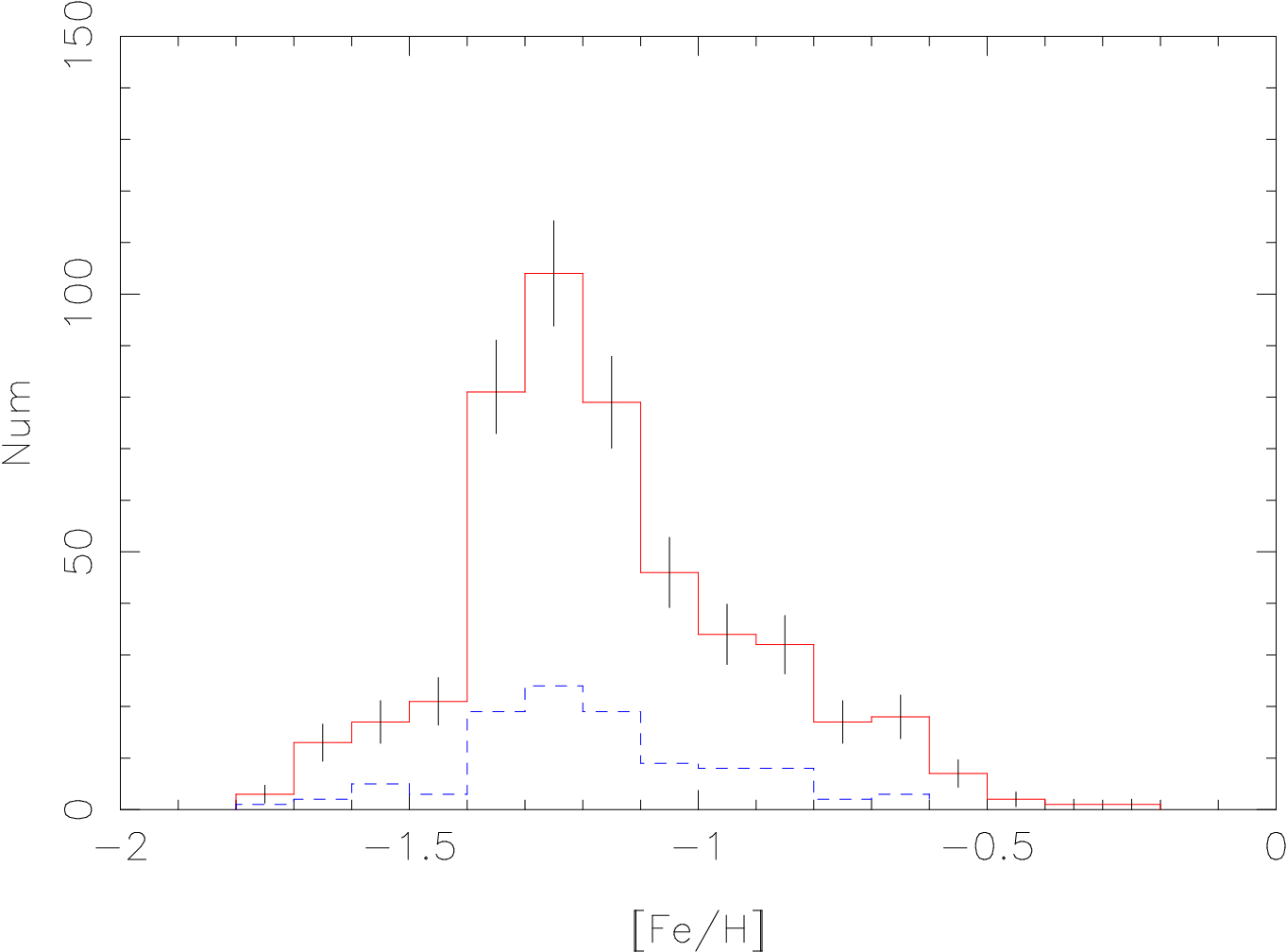}
\end{minipage}\vspace{0.2cm}
\begin{minipage}[t]{0.5\textwidth}
\centering
{\bf }
\end{minipage}
\begin{minipage}[t]{0.5\textwidth}
\centering
{\bf }
\end{minipage}
\begin{minipage}{\textwidth}
\includegraphics[angle=-90,width=75mm]{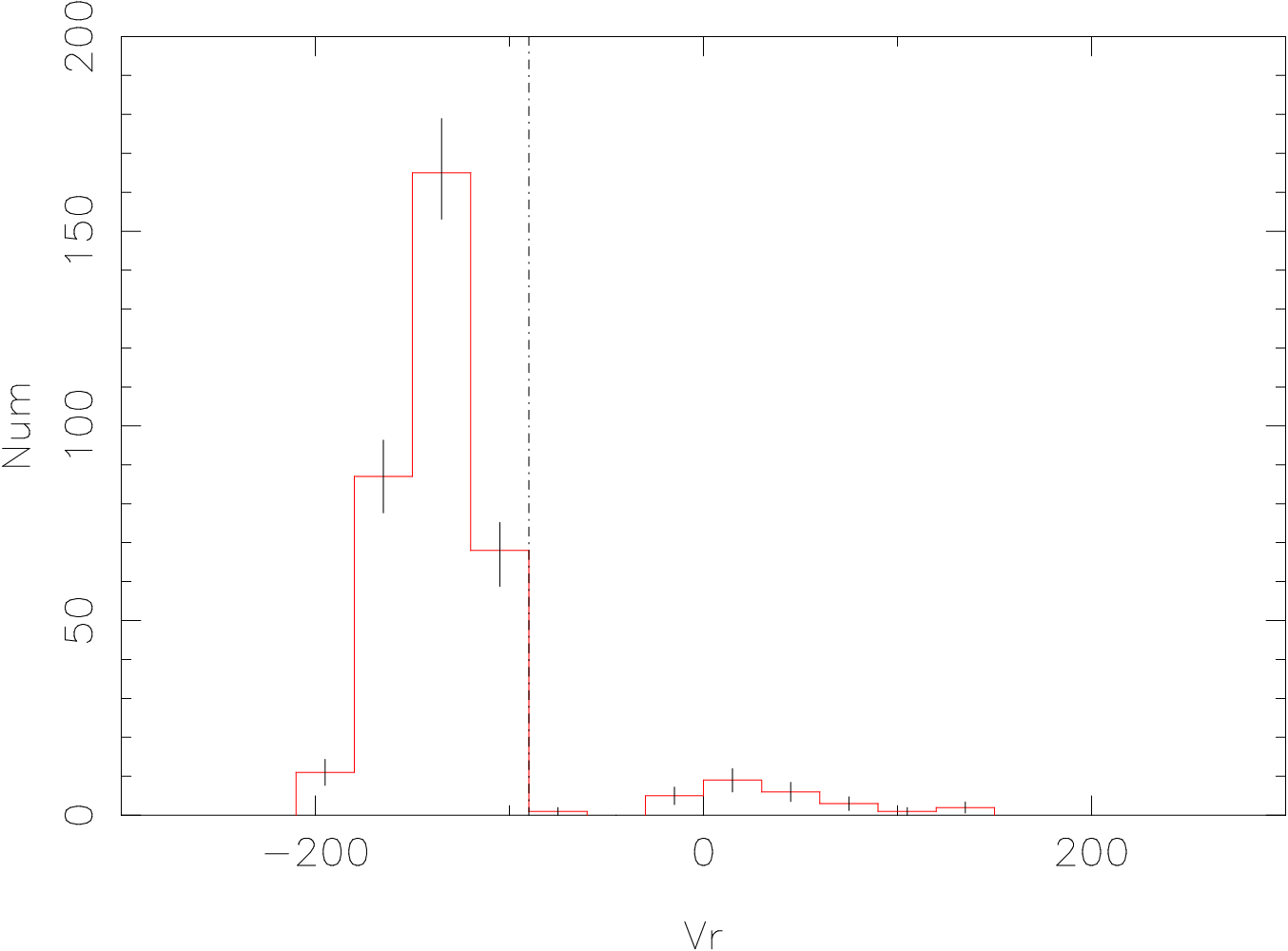}
\hspace*{\fill}
\includegraphics[angle=-90,width=75mm]{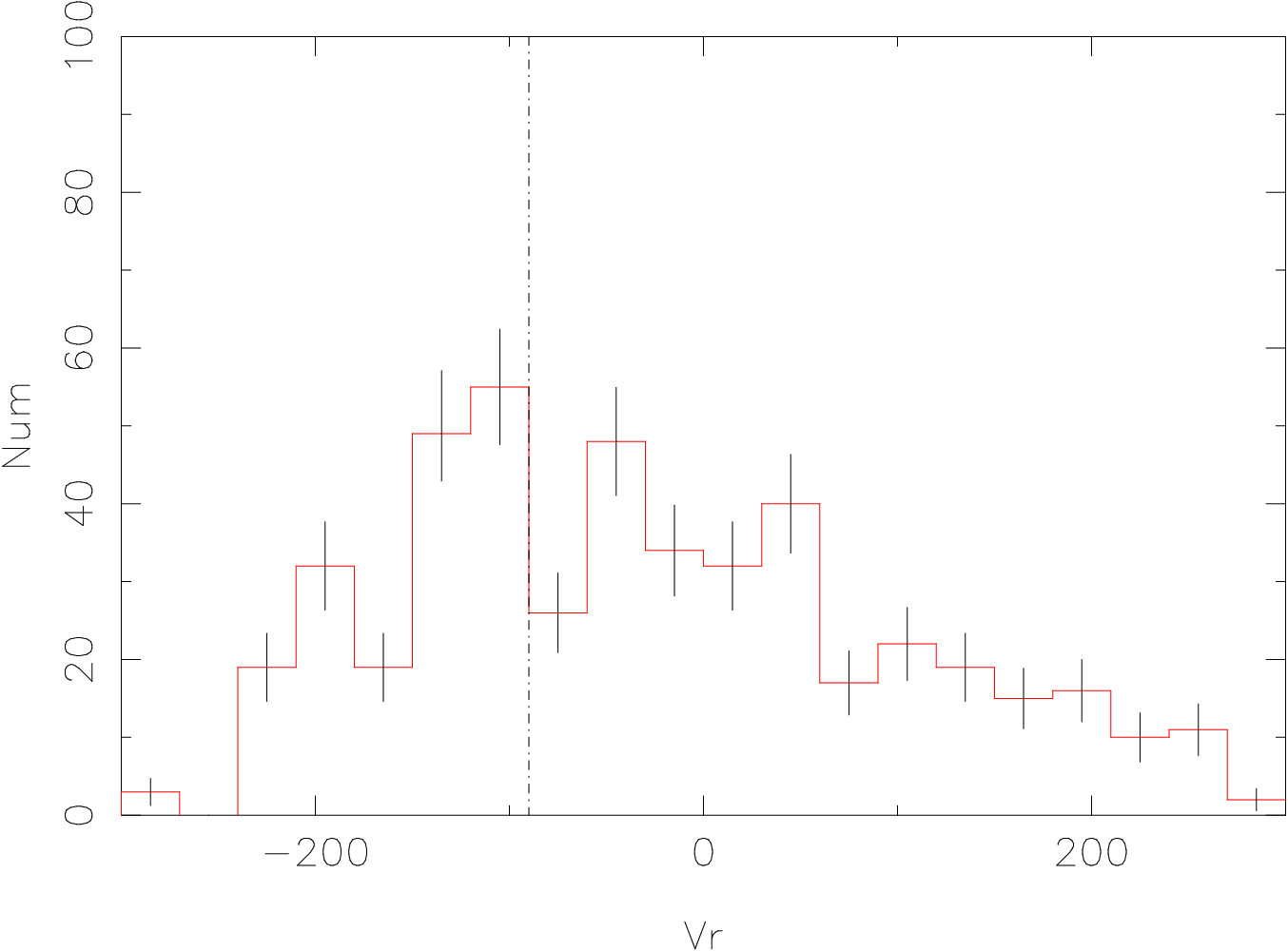}
\end{minipage}\vspace{0.2cm}
\begin{minipage}[t]{0.5\textwidth}
\centering
{\bf }
\end{minipage}
\begin{minipage}[t]{0.5\textwidth}
\centering
{\bf }
\end{minipage}
\begin{minipage}{\textwidth}
\includegraphics[angle=-90,width=75mm]{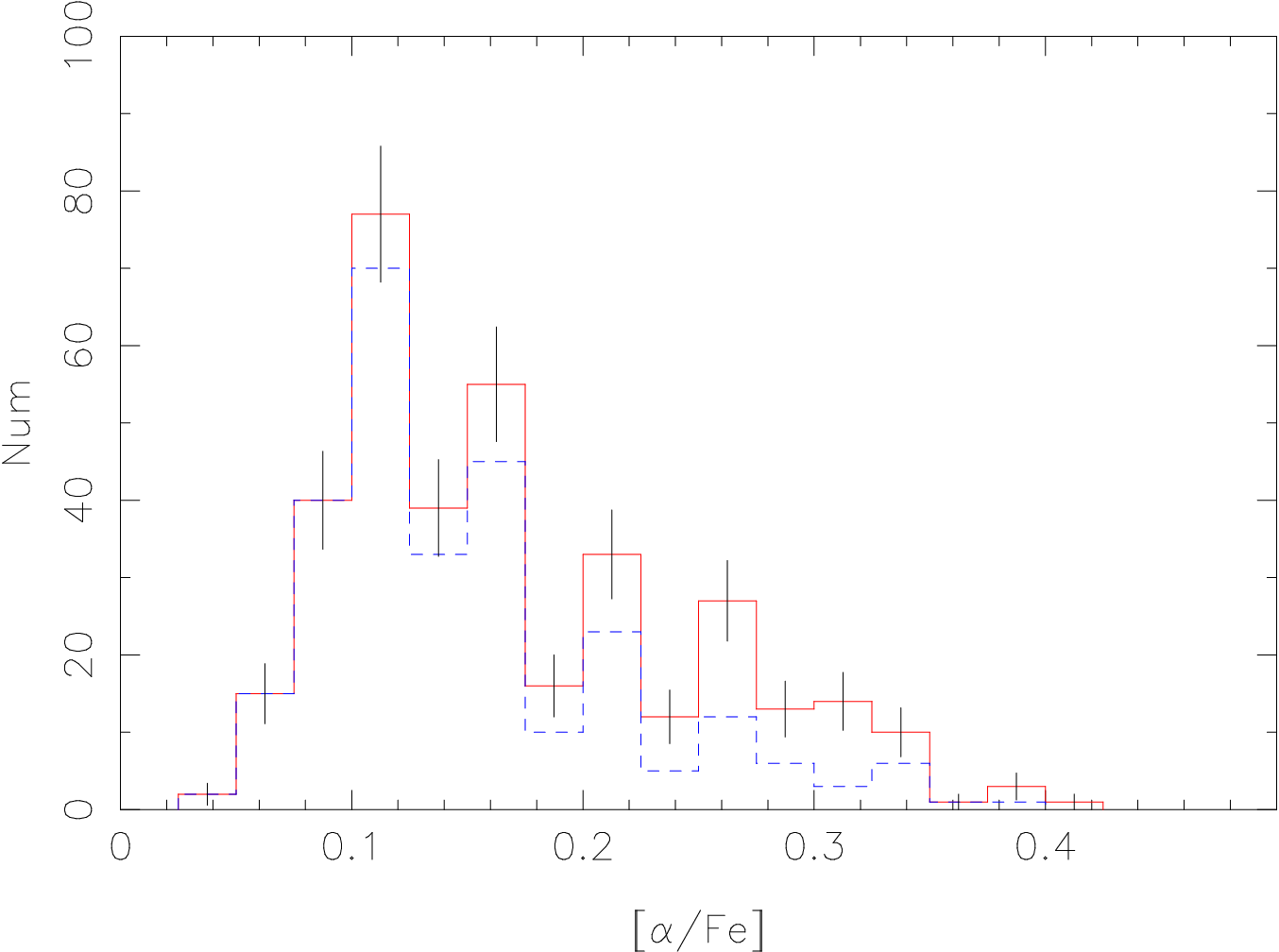}
\hspace*{\fill}
\includegraphics[angle=-90,width=75mm]{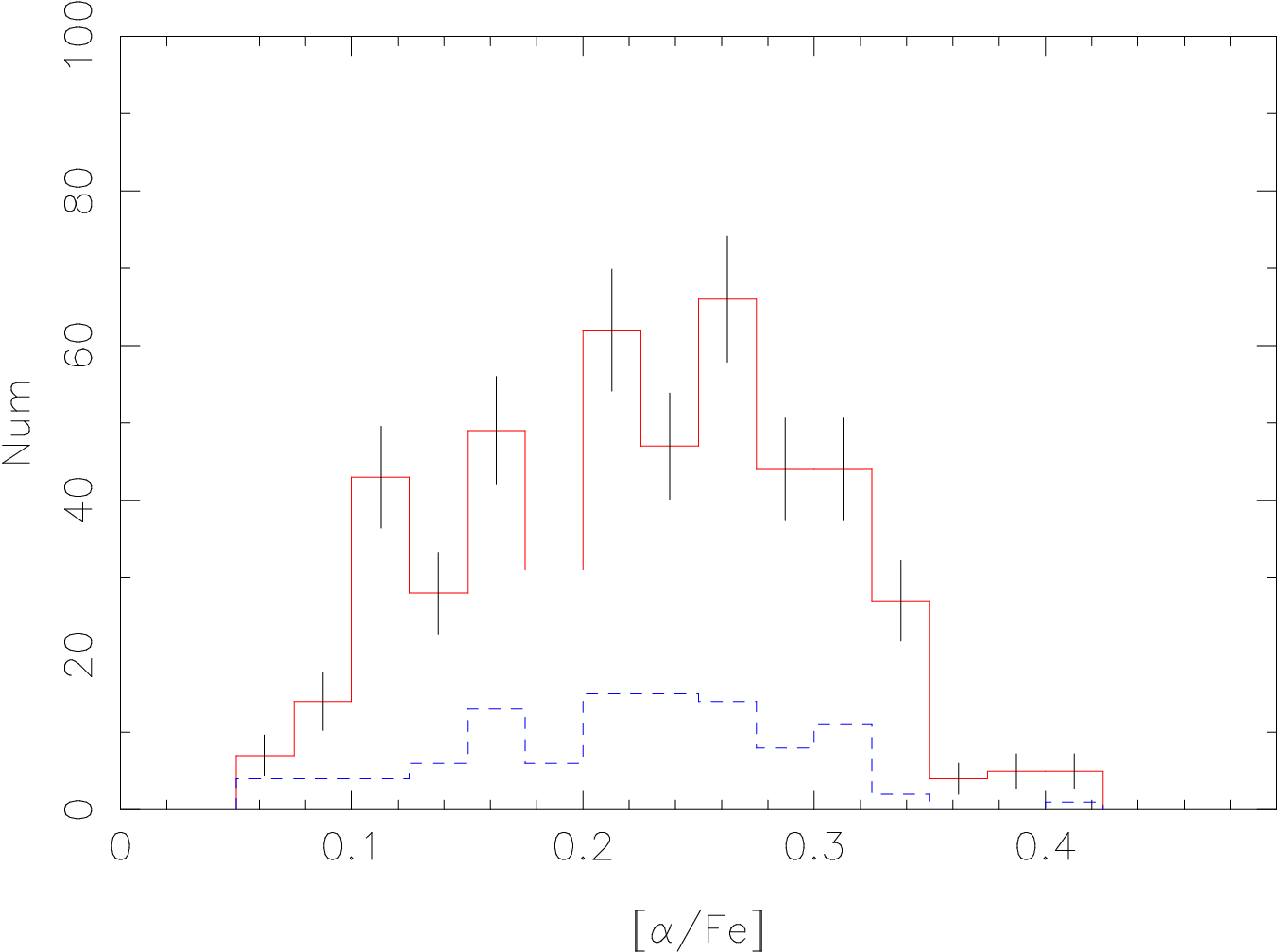}
\end{minipage}\vspace{0.2cm}
\begin{minipage}[t]{0.5\textwidth}
\centering
{\bf }
\end{minipage}
\begin{minipage}[t]{0.5\textwidth}
\centering
{\bf }
\end{minipage}
\begin{minipage}{\textwidth}
\includegraphics[angle=-90,width=75mm]{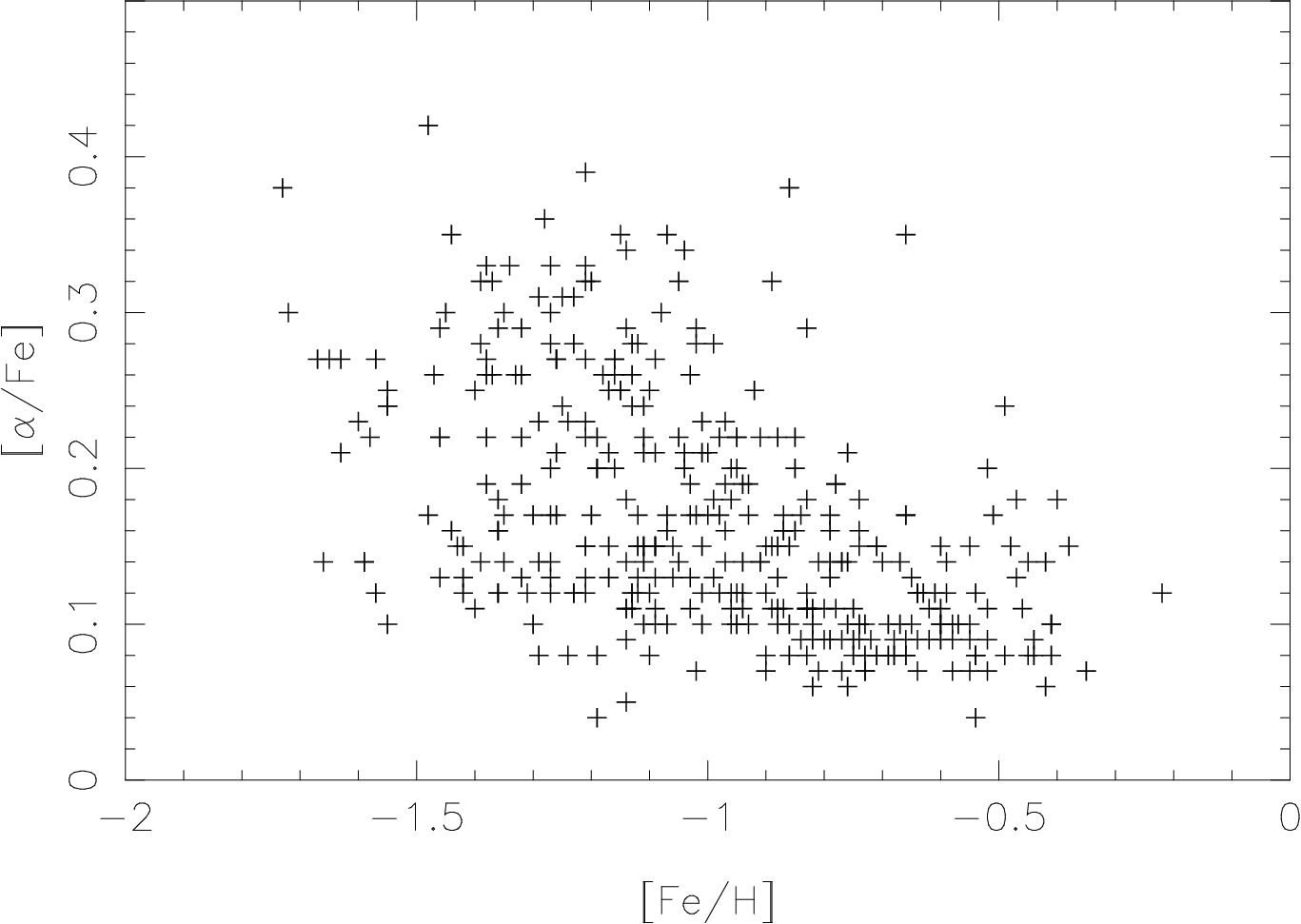}
\hspace*{\fill}
\includegraphics[angle=-90,width=75mm]{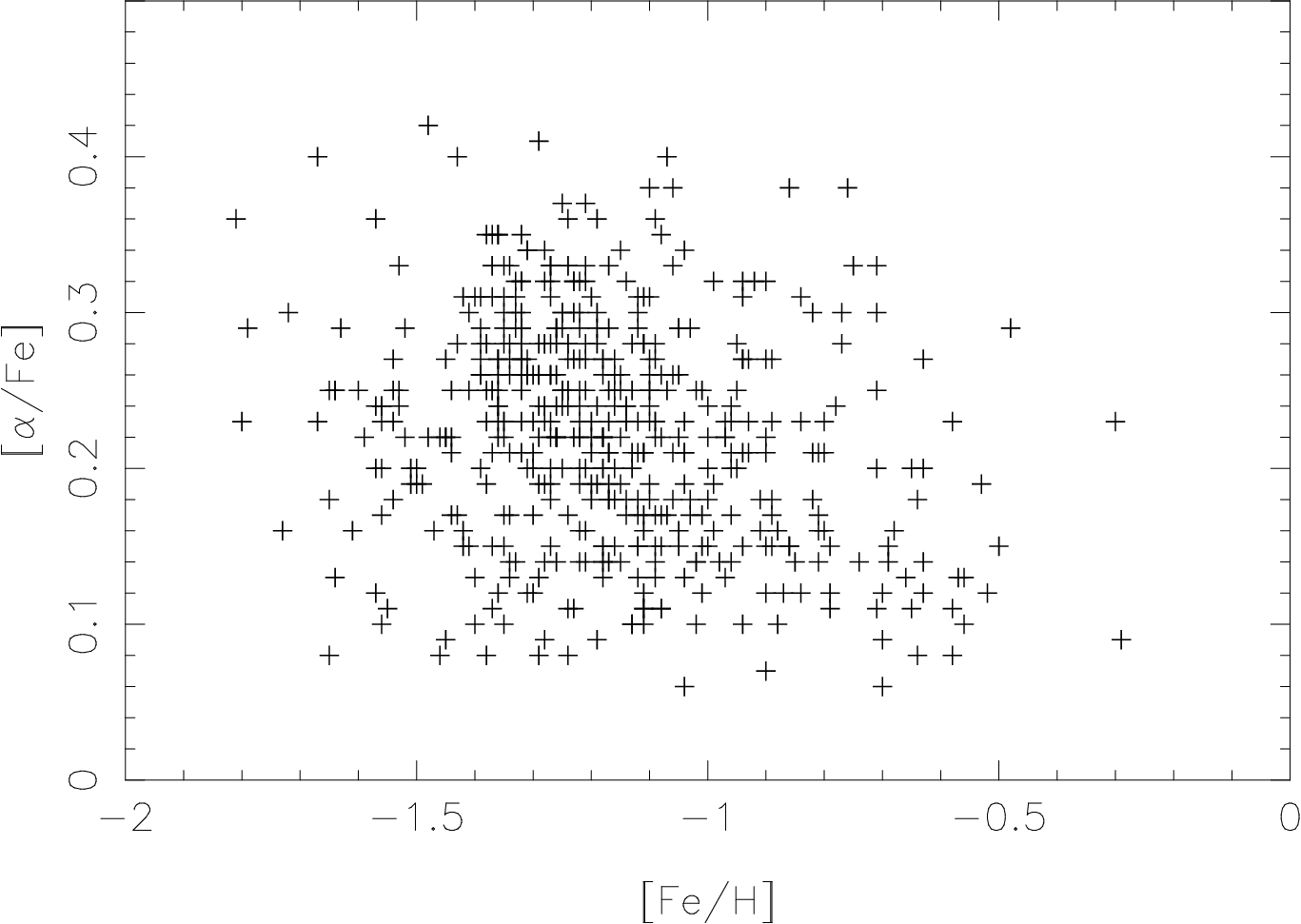}
\end{minipage}
\caption{Same as Figure~\ref{fig:allHisto} but for comparing RHB 
  stars in the trailing arm of the Sgr (358) and Galaxy (476).}
\label{fig:trailing}
\end{figure*}

It is interesting to compare the properties of RHB stars between the
leading and trailing arms of the Sgr tidal tails. Firstly, the
distribution of $V_r$ for the RHB stars have big differences between
the leading and trailing arms (Figures~\ref{fig:leading} -
\ref{fig:trailing}). There are two peaks, -20\,km~s$^{-1}$ and
-100\,km~s$^{-1}$, in the $V_r$ histogram of leading arm stars. The
distribution of the Sgr leading arm stars is similiar with that of the
Galactic RHB stars, which also presents two peaks. Meanwhile, nearly
all the trailing arm stars are centered around one peak near
-140\,km~s$^{-1}$. The dashed line corresponds to low velocity stars
with $V_r$ less than -90\,km~s$^{-1}$ in the $V_r$ histogram, the same
as in Figure~\ref{fig:allHisto}. Again, this difference comes from
the predictions of the \citet{law10} model.  We find that the
metallicity distribution of the stars also has large differences
between the two arms (Figures~\ref{fig:leading} -
\ref{fig:trailing}). The metallicity distribution of RHB stars in the
leading arm is similar to that of the Galactic stars both in the solid
and dashed lines. The metal rich peak is not clear and the metal poor
peak is prominent in the leading arm. The [Fe/H] distribution of the
trailing arm stars has two peaks and the metal rich peak is
significant, as shown in the solid line and even more clearly for low
velocity stars as shown in the dashed line.

From the [$\alpha$/Fe] histograms of Figures~\ref{fig:leading} -
\ref{fig:trailing}, we can see that the distributions of leading arm
RHB stars is also similar with that of Galactic stars both in the
solid and dashed lines, while trailing arm stars show most stars have
lower values of [$\alpha$/Fe], which is different from the Galactic
stars. The properties of the trailing arm RHB stars are more
consistent with the core of the Sgr: [Fe/H] is more metal rich than
that of the Galaxy and [$\alpha$/Fe] is lower than that of Galactic
halo stars. It is unexplained that the leading arm stars do not follow
the chemical history of the Sgr core. Further work is necessary to
investigate the leading arm of the Sgr tidals.

\subsection{The metallicity gradient along the
  leading and trailing arms}

\begin{figure*}[!htbp]
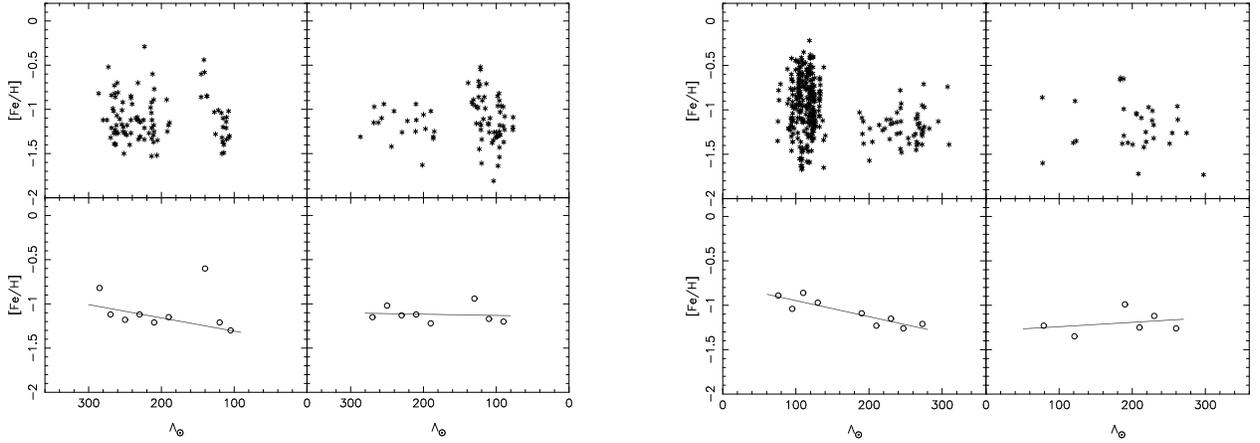

\begin{minipage}{\textwidth}
\centering
\includegraphics[angle=-90,width=75mm]{Figure6A.eps}
\hspace*{\fill}
\includegraphics[angle=-90,width=75mm]{Figure6B.eps}
\end{minipage}
\caption{[Fe/H] as a function of angular distance from the main body
  of Sgr along the leading arm (left panels) and trailing arm (right
  panels). The upper panels show the individual points. In the lower
  panels, the distribution of [Fe/H] is displayed as the median. The
  solid line shows the result of a least-squares linear fit to the
  median data. The metallicity gradient is -(1.5 $\pm$
  0.4)$\times10^{-3}$ dex~degree$^{-1}$ in leading arm 1 and -(1.8 $\pm$
  0.3)$\times10^{-3}$ dex~degree$^{-1}$ in trailing arm 1. The fitted line
  shows that the metallicity is nearly flat in leading arm 2 and 
  trailing arm 2.}
\label{fig:lamda_feh}
\end{figure*}

\begin{figure*}[!htbp]
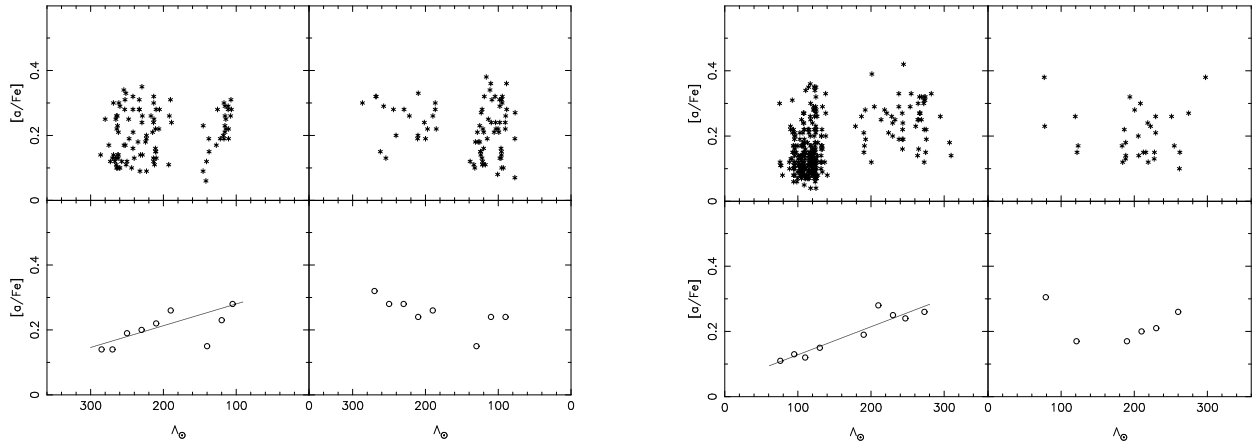

\begin{minipage}{\textwidth}
\centering
\includegraphics[angle=-90,width=75mm]{Figure7A.eps}
\hspace*{\fill}
\includegraphics[angle=-90,width=75mm]{Figure7B.eps}
\end{minipage}
\caption{[$\alpha$/Fe] as a
  function of angular distance from the main body of Sgr along the
  leading arm (left panels) and trailing arm (right panels). The
  [$\alpha$/Fe] gradient is (0.67 $\pm$ 0.15)$\times10^{-3}$ dex~degree$^{-1}$
  in leading arm 1 and (0.86 $\pm$ 0.12) $\times10^{-3}$ dex~degree$^{-1}$ in
  trailing arm 1. There is no obvious trend in leading arm 2 or 
  trailing arm 2 except for the fluctuations of individual points.}
\label{fig:lamda_afe}
\end{figure*}

We would like to see if the metallicity is a function of orbital
longitude along the Sgr leading and trailing tidal streams.
Figures~\ref{fig:lamda_feh} and \ref{fig:lamda_afe} give the
distributions of $\Lambda_\odot$-[Fe/H] and
$\Lambda_\odot$-[$\alpha$/Fe] for RHB stars. There is a metallicity
gradient in trailing arm 1 while there is a lower one in leading arm
1. We find that in the trailing arm, when moving farther from the Sgr
core along arm 1 and then to arm 2, the metallicity shifts to more
metal poor values, which suggests an evolution toward more ancient
stars since metal poor RHB stars must be older than metal rich RHB
stars. This is in agreement with dwarf galaxy formation theories where
the more metal rich core of the galaxy is surrounded by older and more
metal poor stars since it is this outer, older and metal poor
population that will be tidally stripped before the younger, inner
component. Our results also agree with the gradient found by
\citet{cho07} and is similar to Figure~15 of \citet{law10}, which gives
the distribution of $\Lambda_\odot$-[Fe/H]. However, \citet{yan09}
have studied the metallicity of blue horizontal-branch (BHB) stars as 
a function of
$\Lambda_\odot$ and find that there is no significant trend in the BHB
metallicity. It is possible that the BHB stars in \citet{yan09} are
the old and metal poor component of the Sgr, which is not easily
distinguished from the Galactic components with the same properties,
in contrast with our comparison sample.

\subsection{Sgr RHB stars in the bright and faint streams} 
A recent paper by \citet{kop12} suggests the tidal debris 
of the Sgr is actually two separate streams of stars separated 
by $\sim10^\circ$ in the Sgr orbital coordinate system.  Their 
work is an extension of the work of \citet{bel06} who found 
two branches of the leading arm debris in the north Galactic 
cap. The brighter and thicker stream is claimed to have more 
than one stellar population with a large fraction being 
metal-rich. The fainter and thinner stream is said to be 
primarily a single, metal-poor population. No estimate 
of the metallicity of either stream is given by \citet{kop12}, 
but using our RHB sample we can make a qualitative comparison.

\begin{figure*}[!htbp]
\centering
\includegraphics[angle=-90,width=80mm]{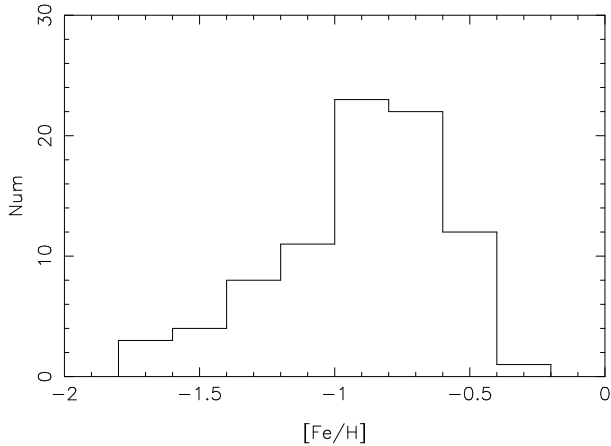}
\caption{The metallicity distribution of stars belonging to the 
  bright stream of \citet{kop12}.  The stars were selected 
  based on the positions and distances given in their Tables 1 
  and 2.}
\label{fig:bright}
\end{figure*}

As we have shown, our Sgr RHB sample is somewhat metal-rich.  
Further, since the brighter stream is also the more metal-rich 
one according to \citet{kop12}, we expect our RHB sample to 
be composed primarily of stars from this stream.  To check if 
this is the case, we separate our sample into stars belonging 
to the bright and faint streams using the positions and distances 
given in Tables 1 and 2 of \citet{kop12}. There are 84 stars 
(80, 2 and 2 from trailing arm, leading arm and overlapping 
group, respectively) in the bright stream and 5 stars (all from 
the trailing arm) in the faint stream.  The metallicity
distribution of the stars in the bright stream is shown in 
Figure~\ref{fig:bright}. As expected, most of the stars are
metal-rich.

We do not have enough stars in the faint stream to make a 
meaningful comparison.  This could be due to a couple of 
factors.  One reason we may not have many stars corresponding 
to the faint stream is because our selection criteria are 
based on the model of \citet{law10}.  This selection may 
preclude these stars simply based on positions and/or 
kinematics. Another possibility is that the metal-poor faint 
stream has little or no RHB component. A more detailed 
description of the two streams is necessary in order to 
distinguish between these two possibilities.

\section{Error Estimate and Model Test}

\subsection{Comparing with Besan\c{c}on model}
In order to estimate the level of contamination from halo RHB stars 
we use the Besan\c{c}on model of the Galaxy \citep{rob03}.  We 
selected stars from all possible Galactic components and applied 
our selection criteria mentioned above.

We find that the possible contribution from the halo in our sample 
varies greatly depending on the area.  In particular, the leading arm 
areas we select will suffer from more contamination than the trailing 
arm areas because of the closer distances, lower velocities, and 
wider spread in velocities, all of which will increase the number of 
expected halo stars.  Further, in the second wraps of the tidal tails 
the model constraints are not as strong and therefore allow for more 
contamination.  Our cleanest sample is that for trailing arm 1 in part 
because of the larger distances, but more importantly, from the narrow 
range of velocities with large negative values.  We also have the 
largest sample of RHB stars in trailing arm 1 so we expect the results 
from this area to be the best and most robust.

The fact that our metallicity gradients for trailing arm 1 and 
leading arm 1 are so similar and agree within errors means that 
contamination in our sample is small and/or has little effect 
on our results.  We also point out that halo contamination is not 
unique to our RHB sample and disentangling the halo component from 
the Sgr component is very difficult since the stars are at the same 
distances and have the same velocities.  Previous work using 
similar selection criteria as our work will also suffer from 
the same problem (such as 
\citet{yan09,mon07,kel10,cor10,carlin12,kop12}).

\subsection{Error Analysis}
In the current models (especially in \citet{law10}), the younger
segments of tidal debris are constrained to match the 2MASS/SDSS
observations while the older segments are regarded as predictions for
where tidal debris might be expected if it extends beyond that which
is currently traced by 2MASS/SDSS. The dynamical 'age' of a particle
in the \citet{law10} model is given by the parameter 'Pcol' where
values of Pcol~$<=$~3 correspond to tidal debris observed by
2MASS/SDSS. In our analysis, we use stars whose 'Pcol' range from 1 to
7 in the model. For more accuracy we could only use the tidal debris
previously observed by 2MASS/SDSS. These parts nearly correspond to
the first wrap of the leading and trailing arms. This would restrict
our results to only arm 1 of the leading and trailing arms. With this
sample, the results become stronger and thus our results are
reliable. In particular, the metallicity and [$\alpha$/Fe] gradients
are only detectable in arm 1 of the leading and trailing arms.

We vary our ranges by 10\% in distance and velocity for the sample
selection to obtain a larger or smaller sample and perform the same
analysis procedure. The results are very similar with the original
ones. This indicates that the sample selection criteria are reasonable
and the results are robust.

\subsection{Distance distributions for stars with
  $\Lambda_\odot<130^\circ$ as a model test}

\begin{figure*}[!htbp]
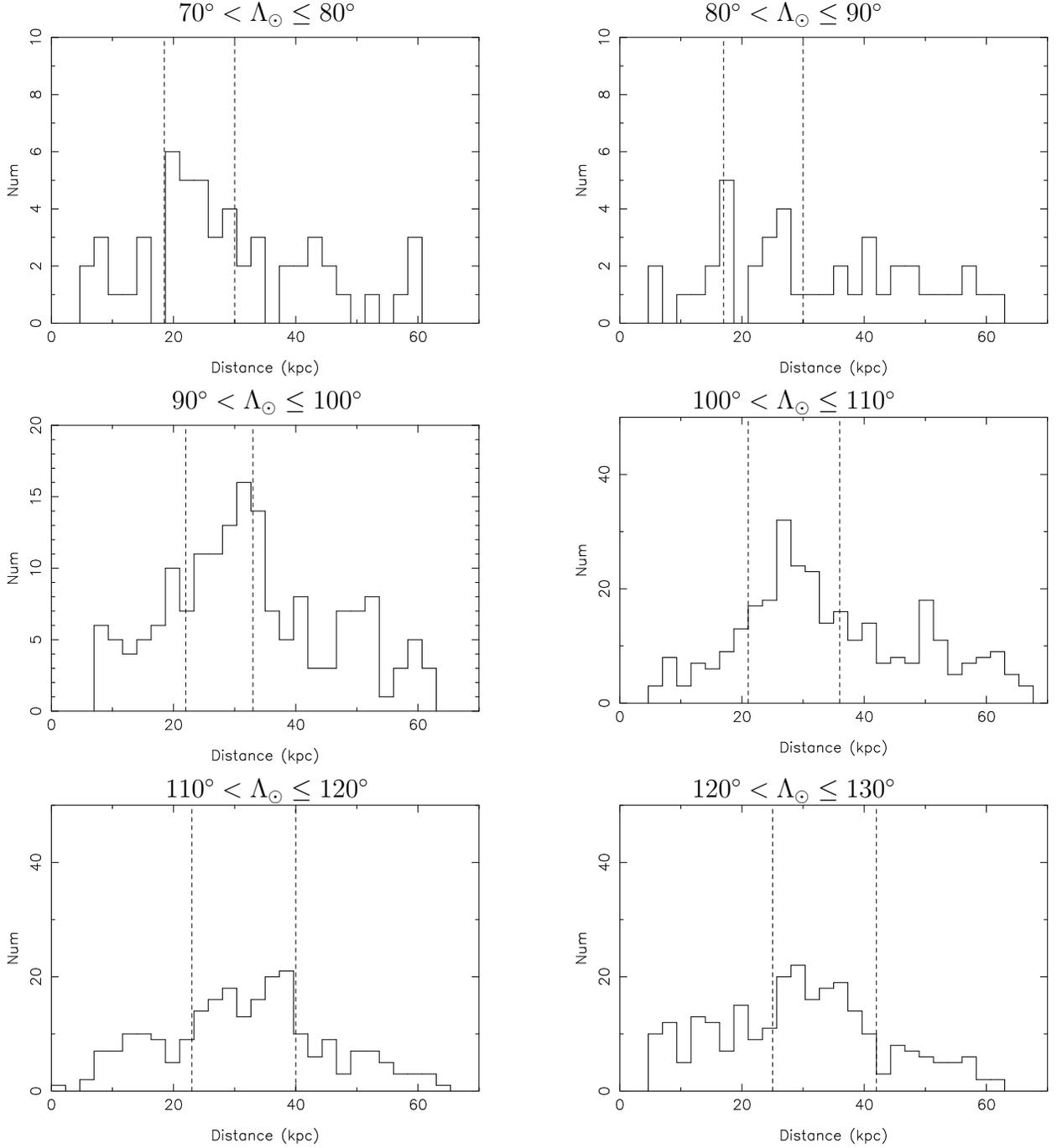

\begin{minipage}[t]{0.5\textwidth}
\centering
{\bf $70^\circ<\Lambda_\odot \le 80^\circ$}
\end{minipage}
\begin{minipage}[t]{0.5\textwidth}
\centering
{\bf $80^\circ<\Lambda_\odot \le 90^\circ$}
\end{minipage}
\begin{minipage}{\textwidth}
\includegraphics[angle=-90,width=75mm]{Figure9A.eps}
\hspace*{\fill}
\includegraphics[angle=-90,width=75mm]{Figure9B.eps}
\end{minipage}\vspace{0.2cm}
\begin{minipage}[t]{0.5\textwidth}
\centering
{\bf $90^\circ<\Lambda_\odot \le 100^\circ$}
\end{minipage}
\begin{minipage}[t]{0.5\textwidth}
\centering
{\bf $100^\circ<\Lambda_\odot \le 110^\circ$}
\end{minipage}
\begin{minipage}{\textwidth}
\includegraphics[angle=-90,width=75mm]{Figure9C.eps}
\hspace*{\fill}
\includegraphics[angle=-90,width=75mm]{Figure9D.eps}
\end{minipage}\vspace{0.2cm}
\begin{minipage}[t]{0.5\textwidth}
\centering
{\bf $110^\circ<\Lambda_\odot \le 120^\circ$}
\end{minipage}
\begin{minipage}[t]{0.5\textwidth}
\centering
{\bf $120^\circ<\Lambda_\odot \le 130^\circ$}
\end{minipage}
\begin{minipage}{\textwidth}
\includegraphics[angle=-90,width=75mm]{Figure9E.eps}
\hspace*{\fill}
\includegraphics[angle=-90,width=75mm]{Figure9F.eps}
\end{minipage}
\caption{The distance distribution of RHB stars for 
  $\Lambda_\odot<130^\circ$. Each panel shows the stars in a $10^\circ$
  bin. The dashed lines show the range of model distances in trailing arm 1.}
\label{fig:trailing1bins}
\end{figure*}

In our sample there are a significant number of stars in trailing arm
1 with $\Lambda_\odot<130^\circ$, but not enough stars for good
statistics in other areas. We thus investigate the distance
distributions of our RHB stars and compare them with model predictions
since we expect that Galactic stars have a broad distribution and
there should be an overdensity when the Sgr stream passes through the
Galactic field. Figure~\ref{fig:trailing1bins} shows the distance
distributions of RHB stars and dashed lines show the distance range
given in the model for Sgr trailing arm 1 for
$\Lambda_\odot<130^\circ$ with a bin width of $10^\circ$.  One sees
that almost all bins show distance peaks within the model predicted
ranges despite the significant selection effect of the SDSS
spectroscopic survey. It seems that the distance prediction in the
\citet{law10} model is correct and our sample selection of RHB stars
based on this model is reasonable.

\subsection{The velocity dispersion at
  $88^\circ<\Lambda_\odot<112^\circ$ as a model test}

\begin{figure*}[!htbp]
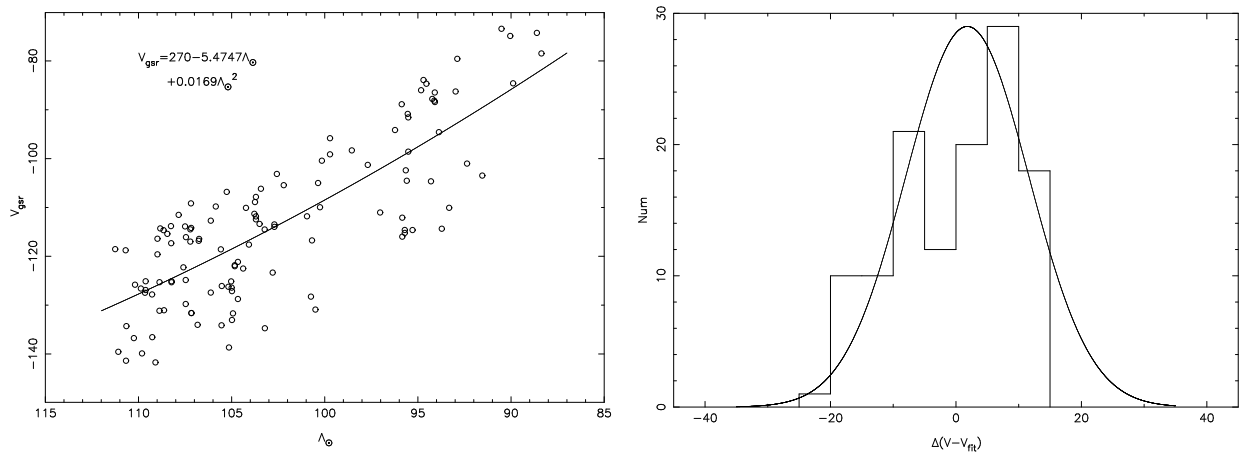

\begin{minipage}{0.5\textwidth}
\centering
\includegraphics[angle=-90,width=80mm]{Figure10A.eps}
\end{minipage}
\begin{minipage}{0.5\textwidth}
\centering
\includegraphics[angle=-90,width=80mm]{Figure10B.eps}
\end{minipage}
\caption{Left panel: radial velocity as a function of longitude of the
  Sgr orbital plane for our RHB stars. The polynomial fit to the
  distribution is also plotted. Right panel: distribution of the
  differences between the polynomial fit and the data. A Gaussian fit
  is also shown.}
\label{fig:velo}
\end{figure*}

Our sample has the largest number of stars at
$88^\circ<\Lambda_\odot<112^\circ$, which covers a similar area as
\citet{maj04} and \citet{mon07}. Thus, we investigate the velocity
dispersion for this area so that we can compare our result to these
works and provide a test to the model prediction. In the left panel of
Figure~\ref{fig:velo}, we plot V$_{gsr}$ as a function of the Sgr
longitude $\Lambda_\odot$. The solid line is a polynomial fit to the
data and it describes a characteristic trend of decreasing V$_{gsr}$
with increasing $\Lambda_\odot$ along the Sgr trailing tail, as
already discussed by \citet{maj04} and \citet{mon07}. The fit is for
$88^\circ<\Lambda_\odot<112^\circ$ because for
$\Lambda_\odot<90^\circ$ an increase of the velocity dispersion is
evident (see \citet{maj04} and \citet{mon07}). The right panel shows
residuals of our sample stars with respect to the polynomial fit. The
distribution is fit with a Gaussian of width
$\sigma$=9.808$\pm$1.0~km~s$^{-1}$ using 119 stars. \citet{mon07} give
a velocity dispersion of $\sigma$=8.3$\pm$0.9~km~s$^{-1}$ using 41
stars with high resolution spectroscopy and \citet{maj04} give
$\sigma$=10.4$\pm$1.3~km~s$^{-1}$ for stars with low resolution
spectroscopy. These three values are consistent within errors. The
agreement indicates that the \citet{law10} model prediction is
reasonable, which is what our sample star selection is based on. These
parts of the trailing tail are dynamically colder than the Sgr core,
which has dispersions of 11.17~km~s$^{-1}$ and 11.4~km~s$^{-1}$ in
\citet{mon05}.

\section{Summary}
 
In this paper we present the properties of the metallicity and
$\alpha$-abundance distributions for a large sample of RHB stars belonging 
to the Sgr tidal streams. The Sgr stars have two components in [Fe/H]
while the Galactic stars have a more prominent metal-poor one. 
[$\alpha$/Fe] is lower for the Sgr stars
than for Milky Way stars, especially along the trailing arm. There are
metallicity gradients along the streams of Sgr, with a value of -(1.8
$\pm$ 0.3)$\times10^{-3}$ dex~degree$^{-1}$ in trailing arm 1 and of
-(1.5 $\pm$ 0.4)$\times10^{-3}$ dex~degree$^{-1}$ in leading arm 1. No
significant gradient exists along trailing arm 2 or leading arm 2.
Stars belonging to more ancient wraps of the streams in arm 2 are more
metal-poor.

We test the model and sample selection in four aspects as follows.
First, by comparing with the Besan\c{c}on model of the Galaxy
  we find that contamination from the Galactic halo is small for
  the largest sample of RHB stars in trailing arm 1.
Then we change the selection range for the width of the leading and
trailing arms and find no significant difference. Third we
investigated the distance distribution of RHB stars in trailing arm 1
($\Lambda_\odot<130^\circ$) and the peaks fall within the model
prediction ranges. Fourth we test the velocity dispersion for the Sgr
trailing tail at $88^\circ<\Lambda_\odot<112^\circ$ and found a value
of $\sigma$=9.808$\pm$1.0~km~s$^{-1}$, which is consistent with the
results of \citet{maj04} and \citet{mon07}.

With the upcoming LAMOST spectroscopic survey, we can expect to
analyze RHB stars in the Sgr for an even larger sample and in
different Galactic locations in order to further study the chemical
history of the Sgr galaxy.

\acknowledgments
We thank the referees for their helpful comments which significantly
improved the paper. This work was supported by the National Natural
Science Foundation of China (Grant No.11178013, 11073026, 11150110135
and 10978015), and by the Provincial Natural Science Foundation of
ShanDong (Y2008A08 and ZR2010AM006).

\end{document}